# Mobility Label Based Network:

## Hierarchical Mobility Management and Packet Forwarding Architecture

Oleg Berzin*

*Verizon Communications, Philadelphia, PA, USA, E-Mail: oleg.berzin@verizon.com

***Abstract* – Scalability of the network layer mobility management solution is one of the most important requirements for the mobility control plane. Mobility Label Based Network (MLBN) is a new approach to the network layer mobility management problem that relies solely on MPLS to provide both macro- and micro-mobility for IPv4 and IPv6 mobile hosts and routers. This new approach does not rely on the existing IP mobility management protocols such as Mobile IP and is based on the combination of Multi-Protocol BGP (MP-BGP) and MPLS. In the context of the MLBN the scalable control plane should be capable of efficient Mobility Label distribution while allowing the MPLS based forwarding plane to deliver mobile traffic in an optimal manner. This paper presents a hierarchical mobility management system capable of both macro- and micro-mobility support without the use of Mobile IP and its derivatives and allows scalable Mobility Label distribution and MPLS label stack based packet forwarding in support of optimal traffic delivery between the communicating mobile users.**

## I. Introduction

The Mobility Label Based Network (MLBN) introduced in [1] is the network layer mobility management system that is independent from Mobile IP [3], [10] and is based on MPLS [9] and Multi-Protocol BGP [6], [7]. It does not require Home Agents, Care-of-Addresses and layer 3 based traffic tunneling to enable communications with and between the mobile nodes equipped with fixed IP addresses and residing outside of their home networks. As a matter of fact MLBN architecture does not have a concept of a home network and always treats the registered hosts or routers as mobile nodes.

The main goal of MLBN is to integrate the layer 3 mobility control plane and the MPLS forwarding plane in order to achieve optimal traffic delivery and thus avoid user and network facing performance penalties associated with inefficiencies of the Mobile IP based solutions. This is achieved by using Mobility Labels (overlay MPLS labels) to represent the current location of a mobile user in the network. Mobility Labels are associated with the IP addresses of mobile hosts or IP prefixes served by mobile routers to form Mobility Bindings. Mobility Bindings in turn are distributed using MP-BGP to other network nodes (MPLS LERs) to identify the current location of the mobile user. The traffic delivery is then based on MPLS label stack and follows the optimal network path.

The benefits of MLBN can be summarized as follows:

1. *Elimination of Mobile IP and its physical and logical components* such as Foreign Agent (FA), Home Agent (HA), Care-of-Address (CoA), Collocated-Care-of-Address (CCoA) resulting in the natural integration of the mobile and MPLS transport networks.
2. *Elimination of user and network facing penalties*. For Mobile IPv4 and Mobile IPv6 in bidirectional tunneling mode: elimination of suboptimal routing due to triangular routing and reverse tunneling. For Mobile IPv4 and Mobile IPv6: elimination of HA scalability issues (tunnel management performance, home link congestion, capacity, home agent failures), natural support for mobile node multi-homing and processing load distribution.
3. *Integration of Mobility Control and Forwarding Planes* under the MPLS framework resulting in optimal traffic management.
4. *No requirement for explicit per-mobile prefix Mobility Label Switched Path (LSP) setup, teardown or redirection.* All Mobility LSP's are preconfigured by means of the Label Distribution Protocol (LDP) and exist at the time of network creation providing fully meshed logical connectivity among the nodes of MLBN. To achieve mobility management, only the mapping of mobile prefixes to existing LSPs is required on the subset of MPLS nodes (LERs) and is accomplished by means of Mobility Binding distribution using MP-BGP.
5. *Optimal traffic delivery* for Mobile-to-Mobile and Mobile-to-Fixed communications without additional requirements on the fixed nodes.
6. *Support for both IPv4 and IPv6* under common MPLS-based Control and Forwarding planes.
7. *Support for Mobile Hosts and Mobile Routers.*
8. *Support for Private Mobile Networks* by means of MPLS and MP-BGP.
9. *Ability to leverage Quality-of-Service (QoS) and Traffic Engineering (TE) capabilities of MPLS* for mobile traffic.

This paper is a continuation of work presented in [1]. It specifically addresses the scalability aspects of MLBN related to Mobility Binding distribution and the associated network update process, and introduces support for both macro- and micro-mobility under the same architecture. The increased scalability is achieved by the introduction of the network hierarchy with distinct mobility control and forwarding functions. This paper introduces the following new elements to MLBN as compared to [1]:

1. *Hierarchical network structure, regionalized architecture and corresponding network elements:* Mobility Regions and Areas, Area Label Edge Routers (ALER) and Area Mobility Route Reflectors (AMRR).
2. *New network update procedures* resulting in the scoping of the network update messaging and minimization of the amount of visitor information state changes that need to be maintained by MLBN nodes.
3. *Use of segmented Mobility LSPs* allowing support for micro-mobility while preserving optimal traffic delivery.
4. *New capabilities* such as: survivability and load distribution, mobile node multi-homing and traffic continuity during hand-offs, virtualization for private networking and inter-carrier roaming.

The set of MLBN control plane processes executing in the MPLS LER is referred to as the Mobility Support Function (MSF). A high level operation of MSF is shown in Fig. 1, where upon the MSF Discovery and Registration procedures a Mobility Label is associated with the IP addresses of mobile nodes and distributed throughout the network in the form of Mobility Bindings by using MP-BGP as the control plane signaling protocol. Mobility Label is a second (inner) label in the MPLS stack. It is followed by the infrastructure or top label that is used for the LSP to reach the MLBN node. The MLBN node terminating the LSP identified by the top label will pop this label and read the inner Mobility Label to identify a mobile device or the next LSP to reach the new location of the mobile device. The use of MPLS label stack allows to implement a scalable mobility management hierarchy.

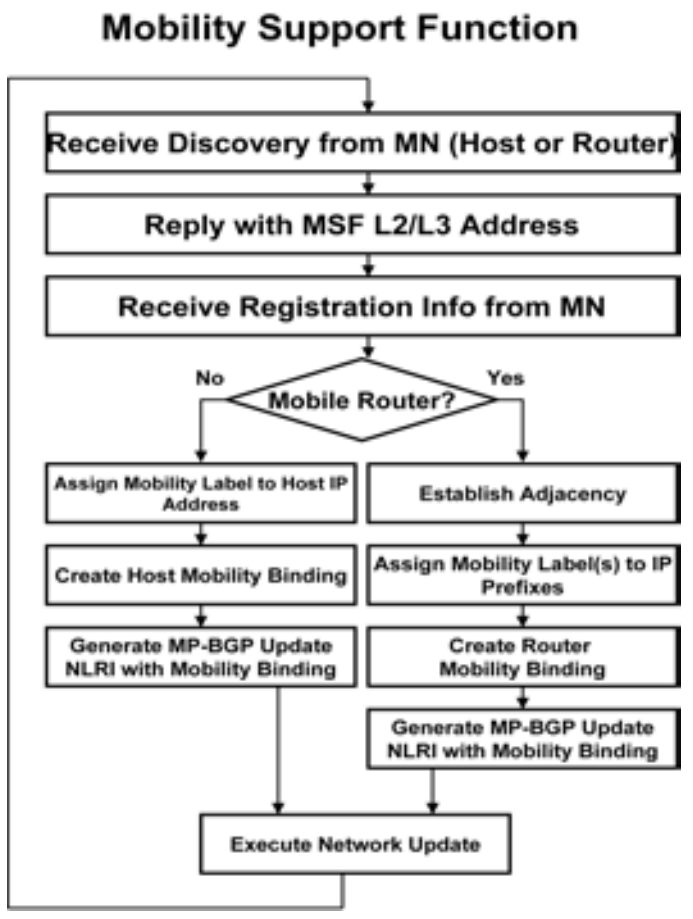

Figure 1. High level operation of the Mobility Support Function (MSF) in the MPLS LER supporting MLBN.

From the architectural perspective MLBN follows the classic MPLS architecture in which the edge LER nodes interface with the Radio Access Network (RAN) by means of the layer 2 grooming network. The network is structured as a collection of the Mobility Regions. Each Mobility Region covers a number of RAN clusters. The mobile nodes register with the serving MSF in the MPLS LER node (see Fig. 2). As mobiles move from one Mobility Region to another different types of hand-offs may be considered.

The hand-offs between the different RANs in the same Mobility Region are referred to as MSF-Local handoffs and do not result in either the new discovery and registration procedures or the network update with the new Mobility Label information. This type of the mobile user movement is also referred to as micro-mobility. When a mobile node moves from one radio cluster to another the MSF tracks this movement and updates the local associations with the new logical layer 3 interface identifier.

The hand-offs between the Mobility Regions are referred to as the Inter-MSF hand-offs. In the hierarchical MLBN Mobility Regions are grouped into Mobility Areas. The hand-off between Mobility Regions within the same Mobility Area is also an example of micro-mobility as no network update outside the area is required. The hand-offs between Mobility Areas are handled by the inter-area update procedures and represent the macro-mobility management.

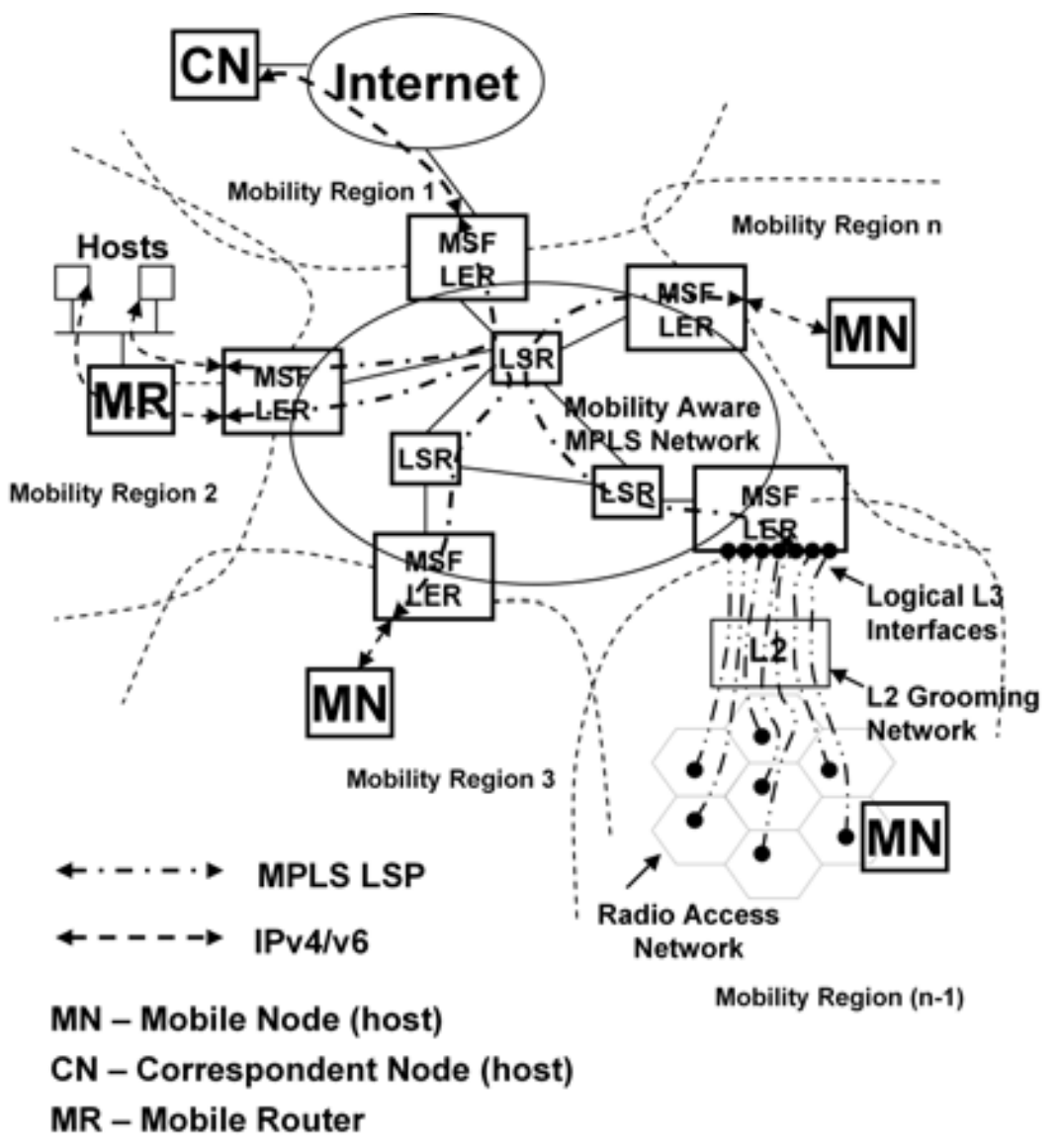

Figure 2. High level architecture of the Mobility Label Based Network.

This paper describes a hierarchical mobility label distribution scheme for the Mobility Label Based Network as well as the corresponding architecture targeted to increase the scalability of the overall solution. The rest of this paper is organized as follows. Section II provides analysis of the existing work. Section III describes the scalability aspects. Section IV introduces the hierarchical mobility management system and describes the packet forwarding architecture. Section V provides additional architectural considerations. Section VI describes various applications for Mobility Labels. Section VII highlights the user facing benefits of the overall solution and Section VIII offers the conclusion.

## II. Related Work

### A. Mobile IP Macro-mobility

*Mobile IPv4*: MIPv4 [3] provides macro mobility management for mobile hosts using IPv4. The main entities in MIPv4 are the Mobile Node (MN), the Correspondent Node (CN), the Home Agent (HA) – the router that owns the original home sub-net, the Foreign Agent (FA) – the router that owns the sub-net to which the MN has moved, and the Care-of-Address (CoA) – the IP address that belongs to the FA and that is used to represent the MN while it is located in the foreign sub-net. The basic operation of MIPv4 requires traffic tunneling between FA and HA (IP-IP or GRE [5]) and results in triangular routing when traffic from CN to MN needs to first go through HA and then use the tunnel from HA to FA to reach the MN. From MN to CN the traffic may follow directly. However, due to the requirement that the source IP address of the packets sent by MN is its Home Address (which is topologically incorrect in the foreign sub-net) the direct MN to CN path may be impossible due to the ingress packet filtering [11]. To overcome this reverse tunneling is used – MN sends packets to FA, FA tunnels packets to HA and HA sends data to CN. Thus with reverse tunneling basic MIPv4 suffers from bi-directional suboptimal routing.

*Mobile IPv4 with Route Optimization*: ROMIPv4 [16] allows CN to send packets directly to MN without going through HA-FA tunnel thus eliminating triangular routing. ROMIPv4 however imposes significant requirements on CN (which is any IPv4 host on the Internet) that make it an unlikely deployment choice for practical applications. Specifically, CN is required to support MIPv4 binding processing as well as to use tunneling to communicate with MN. In addition, ROMIPv4 requires that MN registers and authenticates with CN (just as it does with HA) and thus poses a problem of distribution and management of relevant security information to every CN MN communicates with. ROMIPv4 also imposes additional complexity and processing load on HA that is now required to keep track of CNs and update them with new binding information as MNs move about.

*Mobile IPv4 Network Mobility:* NEMOv4 [17] is a set of extensions to MIPv4 that allow a mobile router (MR) to register its LAN subnets during MIPv4 registration process. NEMOv4 may operate through a MIPv4 FA or in a Collocated Care-of-Address (CCOA) mode. Using the CCOA mode the mobile router registers its CCOA, its Home Address and the LAN prefixes directly with the MIPv4 HA. The mobile router then establishes a direct MR-HA tunnel and the HA is responsible for forwarding the traffic destined to the LAN devices attached to the mobile router through the tunnel identified by the registered CCOA. NEMOv4 does not support route optimization and is subject to triangular routing or bi-directional tunneling resulting in suboptimal traffic routing. In addition NEMOv4 imposes increased load on HA which has to maintain direct MR-HA tunnels (as opposed to the FA-HA tunnels that may carry multiple MIPv4 sessions).

*Mobile IPv6:* MIPv6 [10] provides macro-mobility support for IPv6. It improves MIPv4 by eliminating the need for FA and use of IPv6 Link Local (LLOC) address instead of CoA. MIPv6 allows MN to use its CoA as a source IP address in all packets it sends thus overcoming the ingress packet filtering issue. MIPv6 provides direct support for route optimization by allowing MN to register itself with CN and update it with its mobility bindings. However just like in ROMIPv4 route optimization for MIPv6 requires that CN (any node on the Internet) support MIPv6 and special IPv6 extensions such as routing header and destination option. In addition, route optimization requires a separate return routability procedure executed on both MN and CN and partly via HA in order to ensure that the packets are sent to a correct MN by CN. These additional requirements on CN are fairly significant and may be considered as obstacles in implementing MIPv6 with route optimization. Another mode of operation for MIPv6 is *bi-directional tunneling* in which CN does not need to support MIPv6 and all traffic is tunneled through HA (using IPv6 Generic Packet Tunneling) in both directions via suboptimal routing path (just like in MIPv4). In addition, elimination of FA results in increased load on HA that now needs to manage individual security associations and tunnels for every MN.

*Mobile IPv6 Network Mobility:* NEMOv6 [18] is part of MIPv6 and enables a stationary or a mobile router to register with HA, receive a home address and register the network prefixes it serves with HA. The mobile router establishes a bi-directional tunnel (using IPv6 Generic Packet Tunneling) with it's HA. The HA binds the network prefixes it receives from the mobile router to the router's care-of-address reachable via a tunnel, and advertises the prefixes into the serving IP network. When packets are sent to devices connected to a mobile router and residing in the registered prefixes the IP network routes the packets to the HA and the HA tunnels the packets to the mobile router which forwards them to the destination. Just like in NEMOv4 traffic for the destinations served by a mobile router has to use a suboptimal path via HA.

### B. Mobile IP Micro-mobility

*Mobile IPv4 Regional Registration:* RRMIPv4 [19] aims to minimize the frequency of HA re-registrations due to MN movements. It proposes a hierarchical FA structure in which Regional FA (RFA) and Gateway FA (GFA) act as proxy HA systems by relaying their addresses as CoA addresses for MN. MN registers with RFA and while it changes its location within the serving area of that RFA no HA re-registration is required. While RRMIPv4 provides micro-mobility in a sense that it hides MN movements within a geographical region from HA it does not address optimal routing. As a matter of fact it introduces its own suboptimal routing structure rooted at a given RFA or GFA as the traffic must visit these nodes to be tunneled to MN. RRMIPv4 is still subject to triangular routing, ingress packet filtering and bi-directional tunneling.

*Hierarchical Mobile IPv6:* HMIPv6 [20] uses Mobility Anchor Point (MAP) as a local HA in order to minimize the frequency of MN re-registrations to the real HA thus hiding the MN movements within the service area of a given MAP from HA. MN obtains two CoA addresses LCOA as in MIPv6 and RCOA (Regional COA – belongs to a MAP). MN registers its LCOA with a MAP and RCOA with its HA. MAP is responsible for tunneling packets destined to the MN's Home Address to its LCOA. As in RRMIPv4 HMIPv6 has its own suboptimal routing structure rooted at a given MAP and does not provide optimal traffic routing (at least in the bi-directional tunneling mode). MIPv6 route optimization applied in the HMIPv6 environment still requires MIPv6 and special IPv6 header/options support on a CN.

### *C. MPLS Micro-mobility*

The MPLS Micro-Mobility [11]-[15], [21] combines Mobile IP architecture and MPLS traffic forwarding to implement mobility support solutions in which Mobile IP is used for macro mobility and MPLS is used to support micro-mobility in the part of the network that interfaces with mobile hosts – MPLS domain. Micro-mobility reflects mobile host movements that can be handled without the re-registration with the Mobile IP HA.

In [21] the mobile host registers with a hierarchical set of special MPLS Label Edge Routers referred to as Label Edge Mobility Agents (LEMA). The LEMA at the top of the hierarchical set is registered with the Mobile IP HA as the FA for the MN. A mobile host receives advertisements from the Access Routers (AR) containing the addresses of a subset of LEMAs and their relationship in the LEMA hierarchy. Mobile hosts chooses a set of LEMAs to register with and the LEMA at the top of the registration tree registers the mobile host with the HA. HA tunnels all packets from CN to MN to the top level LEMA as in regular Mobile IP. Once packets are received and de-encapsulated from the tunnel at the LEMA, the packets are sent on the MPLS LSP to the network location of the MN using the MPLS labels assigned to the MN's IP address as the result of the registration process. As the MN moves to new locations, the hand-off procedures are invoked that start with the MN requesting the hand-off and the LEMA(s) performing the set of signaling steps resulting in the redirection of the MPLS LSP from the old serving LEMA to the new serving LEMA. If the MN movement results in a condition in which the old top level LEMA can no longer serve MN, MN re-registers with the new hierarchical set of LEMA(s) and the top level LEMA is registered as the FA with the Mobile IP HA. Although in [21] MPLS Micro-Mobility makes use of the MPLS traffic forwarding it still is an extension of Mobile IP and requires mobile hosts to implement a complex logic involving a set of registrations such as a registration to the local serving Access Router, registrations to the hierarchy of LEMA(s) and the Mobile IP registration to the HA. The scheme in [21] requires heavy use of signaling starting from the need at the MN to understand the LEMA serving network hierarchy and maintain the registration chain of the serving set of LEMA(s), the requirements for the LSP redirection during the hand-off and finally a requirement for every LEMA node in the communications path to participate in the mobility signaling. The scheme in [21] does not address optimal traffic routing as the overlay LEMA network introduces its own suboptimal tree rooted at the lowest LEMA in the hierarchy with respect to a given MN. In order to provide for truly optimal traffic routing within the MPLS domain every node in the domain would have to be a LEMA (including the AR). In addition, it does not offer support for mobile routers.

In [15] a two-level hierarchy in the MPLS domain is proposed where a LER/FA (Label Edge Router/Foreign Agent) is placed at the edge of the network and the LERG (Label Edge Router Gateway) is connecting the MPLS domain to the Core IP network that contains Mobile IP HA. LERG acts as a proxy HA for MN and registers its own address with HA on behalf of MN. The rest of the MPLS nodes (LSR – Label Switch Routers that are other than the LER/FA and LERG) in the domain do not need to understand Mobile IP. However these nodes do need to participate in mobility management on a per-mobile device basis since all MN movements within the MPLS domain result in the establishment of new or redirection of existing LSPs between the serving LER/FA and LERG by means of the RSVP signaling protocol (explicit LSP setup), thus affecting the scalability of the proposal. In the Fast Handoff mode [15] requires a heavy use of signaling to setup new LSPs from LERG to the new anticipated LER/FA nodes neighboring the current serving LER/FA for MN which further impacts scalability. In the Forwarding Chain mode a set of LER/FAs is advertised to MN and MN is required to choose the LER/FAs from the set it has to register with. In addition, following the movements of MN a series of LSP redirections (from old LER/FA to new LER/FA) by means of explicit LSP setup using RSVP is executed to support traffic continuity during handoffs. As in all other micro-mobility proposals [15] does not directly address optimal traffic delivery as it introduces its own suboptimal routing structure rooted at a given LERG especially for MN-MN communications.

In [12] a concept of a Micro-cell Mobile MPLS (MM-MPLS) is introduced in order to address the issues of suboptimal traffic routing within the MPLS domain. MM-MPLS is an evolution of the Hierarchical Mobile MPLS (as in [15]) in which intermediate LSRs between the LER/FA and LERG (or MPLS-aware FDA) are required to keep track of the MN home addresses and the associated MPLS labels (distributed via RSVP). This allows to introduce a concept of a crossover LSR to handle the handoffs from one micro-mobile domain to another without having to redirect LSPs from old LER/FA to new LER/FA. When MN registers with new LER/FA a RSVP signaling message is sent to LERG. At that time an intermediate LSR that has the information about the old LSP for the same MN home address intercepts the RSVP message and redirects the old LSP to the new LER/FA thus avoiding the use of the old LSP and the associated suboptimal path. This however, comes at the expense of significant added complexity and processing in the MPLS domain where every LSR is required to maintain the state information for every MN and explicit LSP signaling on a per-MN basis is required to handle both the initial MN registrations and the subsequent handoffs.

In [13] the concept of MPLS micro-mobility domain is expanded to include the Radio Access Network (RAN) by requiring RAN Base Stations (BS) to act as IP/MPLS nodes (LER) as well as Mobile IP FA. A Gateway (GW) is used to connect the MPLS/RAN to other IP networks. This gateway acts as both MPLS LER and Mobile IP HA. The authors in [13] recognize scalability issues related to the per-MN LSP management via RSVP and propose the use of LDP (Label Distribution Protocol) to establish any-to-any pre-constructed LSP connectivity within the MPLS domain. Since the BS nodes and the GW are all MPLS-aware the Mobile IP signaling is integrated with MPLS and is used to map the mobile nodes' home addresses to existing LSPs. This proposal, however, does not address optimal traffic delivery (specifically for MN-to-MN communications) as all LSPs are anchored at the GW node. In addition to the complexities related to integrating IP/MPLS functionality into a RAN BS, the GW in this proposal is a single point of failure and a potential source of congestion as well as other scalability issues related to the large scale MN management.

In [14] an IPv6 specific solution is proposed that introduces MPLS into HMIPv6. The MPLS micro-mobile domain uses RSVP signaling to manage LSPs on a per-MN basis thus requiring all MPLS nodes to participate in mobility management. In the overlay model MPLS is used within the HMIPv6 environment purely for the traffic delivery purposes (instead of IPv6 GTP tunnels). The authors point out that the lack of integration between MPLS and Mobile IP results in protocol inefficiencies requiring removal and reinsertion of MPLS headers for every packet by the Mobile IP entities (MAP, HA). The integrated model avoids this by allowing a Mobile IP entity to directly access the MPLS forwarding base. This proposal however is essentially the same as HMIPv6 but with added MPLS capabilities.

### *D. Summary*

As can be seen from the description above there are numerous proposals for handling macro- and micro-mobility. However, a common requirement for all of them is the use of Mobile IP which even in the case of MIPv6 is not immune from sub-optimal routing. In addition, Mobile IP HA or its derivatives (LERG, LEMA, GW) is still a single entity that provides mobility support and therefore represents a central resource that is subject to survivability, capacity and scalability considerations.

The MPLS-based proposals all concentrate on providing micro-mobility and act as extensions to Mobile IP. The MPLS micro-mobility schemes do not directly address optimal traffic delivery issues (especially in the case of MN-to-MN communications) and in turn raise significant scalability concerns due to the need in most of them for explicit per-MN LSP management requiring every node in the MPLS domain to take part in the mobility management.

Throughout all of the considered related work none of the schemes provides a common control plane for scalable micro- and macro-mobility support for IPv4, IPv6, mobile hosts and mobile routers as well as the associated forwarding plane that is capable of optimal traffic delivery even for MN-to-MN communications.

The hierarchical MLBN proposed in this paper is expected to provide all of the above capabilities by departing from the "adjunct-to-MIP", micro-mobility only solution and providing a distributed all-MPLS-based mobility management system.

## III. Scalability Considerations

The scalability of the network layer mobility management solution depends in a large degree on the scalability of the corresponding control plane and may be viewed as the ability of the control plane protocol and the corresponding underlying architecture to store, process and handle the state changes related to the large amount of geographically distributed mobile devices.

The major factors affecting the scalability include the control plane protocol framework, the underlying signaling and forwarding network architecture as well as the processing capabilities of the network nodes participating in the mobility management.

We explicitly distinguish between the control plane (signaling) architecture and the packet forwarding architecture. Although the two functions may be combined in a single device it is expected that separation of these functions will result in greater scalability (specifically where it is related to the processing power requirements for the signaling function). Both of these architectures may be constructed in a flat (single level of hierarchy) or multi-level (hierarchical) manner. We expect that as with any large scale network solutions the increased scalability may be achieved by designing the hierarchical control plane integrated directly with the hierarchical packet forwarding plane. The mechanisms involved in the design for such a solution are based on the MP-BGP protocol capabilities, architectural hierarchy and regionalization as well as the use of MPLS label stack in which the infrastructure (top) labels identify the MLBN network nodes themselves and the mobility (inner) labels identify the mobile prefixes (belonging to either mobile hosts or served by mobile routers) or the next LSP segment to reach these prefixes.

The signaling protocol proposed for the use in the MLBN is based on MP-BGP and allows for the distribution of the Mobility Labels and the corresponding Mobility Bindings throughout the network in an overlay manner. Mobility Bindings are the association between the mobile prefix, MLBN LER Router-ID (IP address that identifies the LER) and the MPLS label (Mobility Label) used as the inner label in the label stack. Mobility Bindings are distributed using MP-BGP and encoded as discussed in [2] using a specific Network Layer Reachability Information (NLRI) format. The association between the inner label and the outer label is derived from the LER's Router-ID. This implicit association is very important for increased scalability. To clarify this point, consider the following.

It is imperative for scalability reasons that the MPLS LSP setup, teardown and redirection in support of mobility management are not performed by every node in the MPLS network (LER and LSR) following the movements, connects and disconnects of individual mobile devices. The existing MPLS micro-mobility proposals have to deal with this issue by reconciling the tradeoff between the ability to provide for optimal routing (when all or almost all MPLS nodes are aware of mobile prefixes and their LSP associations) and the scalability.

The MLBN architecture avoids this by using LDP driven pre-constructed full logical mesh of node-to-node LSPs that connect Router-IDs of all MLBN nodes. These LSPs are established at the time of network creation and do not change unless the changes in the network infrastructure occur. The FECs (Forwarding Equivalency Class) served by these LSPs are only the individual Router-IDs of the MLBN nodes. These LSPs are identified by the LDP assigned infrastructure labels (used as top labels in MLBN). When mobile devices connect, disconnect or change their location in MLBN, the corresponding Mobility Bindings are distributed to the LER nodes using MP-BGP in an overlay manner transparently to the LSR nodes. Since a Mobility Binding carries the Router-ID of the originating LER the corresponding infrastructure label to reach that LER from any other node in the network already exists in the MPLS forwarding tables. The Mobility Binding in MLBN does not setup an LSP it is used to map the mobile prefix and the associated Mobility Label to the existing LSP identified by the originating LER's Router-ID.

This in turn enables to delegate mobility management to a sub-set of the MPLS network nodes. In a simple and flat design this sub-set would only include the edge nodes of the network (the MPLS LER nodes running MSF) as shown in Fig. 2. Fig. 3 shows a flat architecture with both the control and forwarding plane functions combined and in which the sub-set of nodes performing mobility management consists of all edge nodes of the network and excludes the core (LSR) nodes. Please note that the Router IDs of the LER nodes in Fig. 3 are shown as the solid black dots.

It is important to point out that in Fig. 3 only the LER nodes are logically connected using a full mesh of MP-BGP control sessions between the Router IDs of the nodes (six control sessions). At the forwarding level, there is a full logical mesh (not shown in Fig. 3) of MPLS LSPs connecting the Router ID's of all nodes (fifteen LSPs). The Mobility Bindings are distributed by a MSF in a given source LER node using MP-BGP control sessions (in this case the information is flooded to all other LER nodes).The information in the Mobility Binding is used by the destination LER node to map the IP address or prefix of a mobile host or a mobile router to the existing LSP by means of the communicated Router ID that belongs to the source LER.

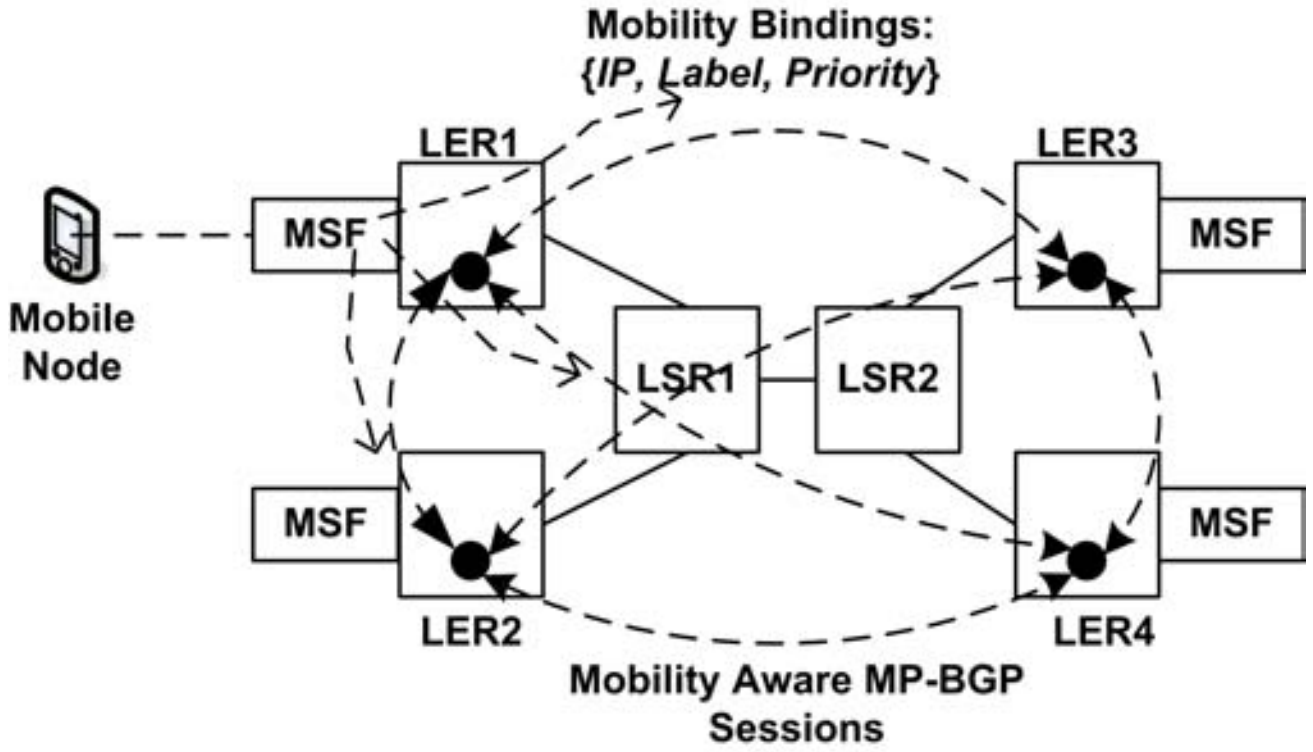

Figure 3. Flat signaling and forwarding architecture with Mobility Binding flooding.

Clearly, this flat architecture cannot be expected to scale due to the need for all of the edge nodes to share the state information (the Mobility Binding Database) and the processing load requirements. However, it is important to note that even in the flat architecture there is room for scalability improvements. Specifically, due to a simple observation that not all edge nodes need to know about all mobile devices at nearly the same time, partitioning of the Mobility Binding information may increase the scalability. For example, if the service used by a mobile device involves only the connectivity from a service provider's network to the Internet, then only the serving MSF and the MSFs residing in the LER nodes connected to the service provider's internet gateways need to share the Mobility Binding information for the mobile device in question. The partitioning mechanism may be based on the existing MP-BGP capabilities such as Route Target extended communities (used in layer 3 MPLS/VPN [4]) or as was presented in [1] on the selective or predictive network update modes (these update modes require modifications to the existing protocol behavior of MP-BGP).

In addition, the improvements in scalability may come from another observation. Namely, it is not necessary to execute the network update process in order to distribute the Mobility Bindings that may not be used for actual communication. In other words the "On-Demand" principle may be followed: "Do not distribute Mobility Bindings unless specifically asked for it". This enables the distribution and updates of the state information confined only to local or regional network nodes performing mobility management. The "On-Demand" principle was introduced in [1] and was referred to as the Hierarchical On-Demand network update mode.

The Hierarchical On-Demand network update mode actually combines three scalability improvement mechanisms: the On-Demand principle, the separation of the control and forwarding plane functions and the control plane hierarchy.

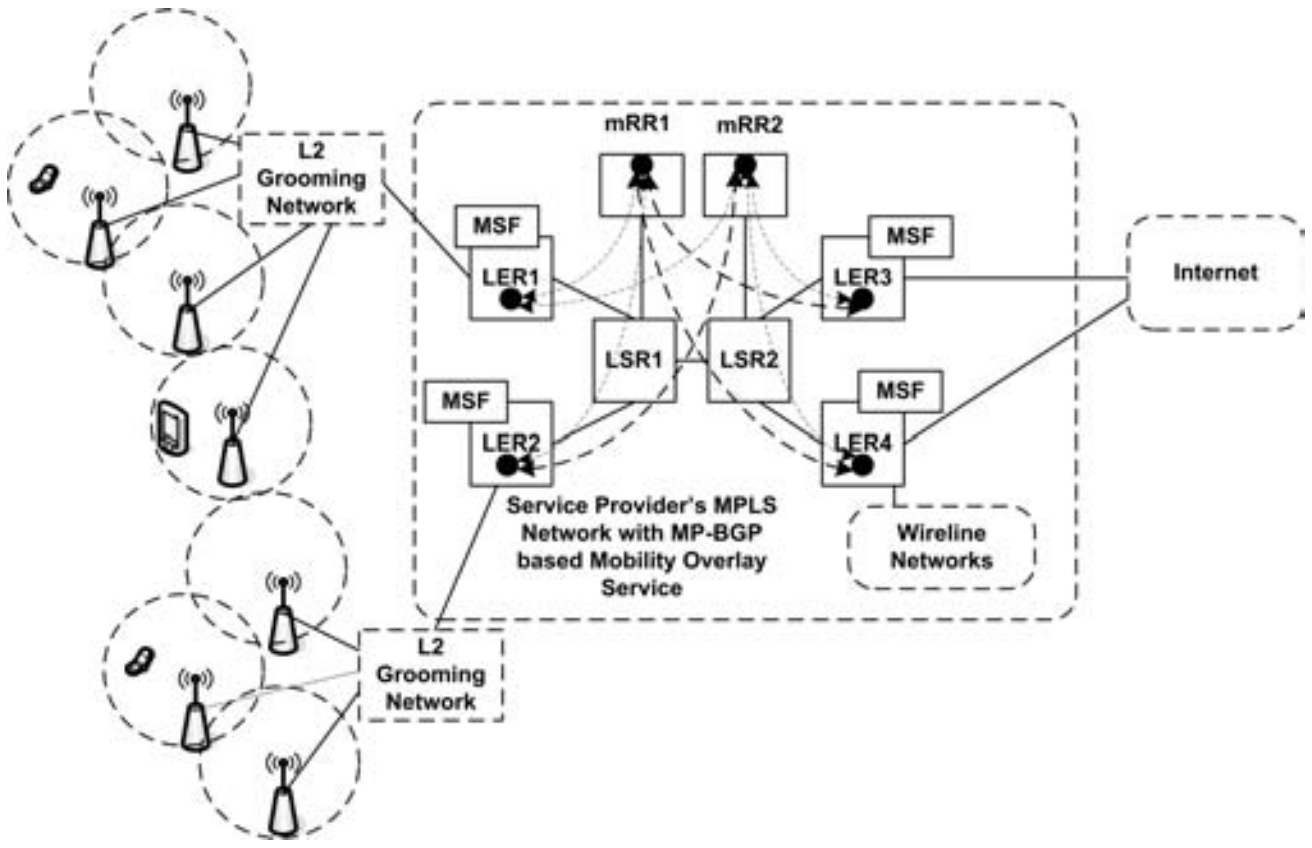

Figure 4. Separation of control and forwarding plane functions and control plane hierarchy.

In Fig. 4, the control plane separation and hierarchy is achieved by introducing another architectural entity – a Mobility Route Reflector (MRR). The MRR peers directly with the edge MSF LER nodes and does not perform packet forwarding functions. The combination of the MRR and the "On-Demand" principle results in the following processing logic. Upon the discovery and registration of the mobile nodes the local serving MSF in the edge nodes assigns the Mobility Label, creates the Mobility Binding and executes the Selective network update to send the information to the MRR. In normal BGP operation the Route Reflector would immediately update all other peering clients with this information. The imposition of the "On-Demand" principle modifies this behavior by requiring the MRR to only store this information and distribute it after an explicit request from another peering node.

At the time when a packet destined to a mobile device arrives at an ingress MSF LER, the LER examines the destination IP address in the packet's header and determines that the address belongs to the mobile address range (this assumption was described in [1]). Assuming that no Mobility Binding information exists at the ingress LER, following the on-demand logic, the LER sends the request for the Mobility Binding information associated with the IP address in question to the MRR. After the MRR replies with the binding, the ingress LER imposes the label stack on the packet (Mobility Label and the infrastructure label associated with the Router ID of the egress LER - the LER that is serving the destination mobile device). The packet is then delivered to the destination LER using MPLS, de-encapsulated to expose the Mobility Label, and then based on the local Mobility Binding Database lookup delivered to the link layer address of the mobile device using the corresponding RA or sent on the next LSP.

In summary, we consider the following protocol and architectural aspects that are expected to increase the scalability of the mobility management scheme: i) separation of the control and forwarding plane functions, ii) on-demand mobility information distribution, iii) hierarchical control plane architecture, iv) hierarchical packet forwarding plane and v) mapping of mobile prefixes to existing MPLS LSPs. The MP-BGP signaling messaging and the use of the stacked MPLS labels bring all of the above elements together into the Hierarchical Mobility Management and Forwarding System.

## IV. Hierarchical Mobility Label Based Network

The hierarchical mobility management system presented here is based on the regionalized network architecture. The following definitions of the architectural entities are used:

*Label Edge Router (LER)* – an edge node in MLBN. LER connects to RAN using L2 grooming network. Each RAN may be terminated at a logical L3 interface of the LER which represents an IP sub-net. LER implements a Mobility Support Function (MSF) and peers with Area MRR using MP-BGP.

*Router-ID (RID)* – an IP address that uniquely identifies a node in MLBN. All Router-IDs in MLBN are reachable via pre-configured LSPs using infrastructure (top) MPLS labels managed by LDP.

*Mobile Prefix* – an IPv4 or IPv6 address of a mobile host or a network prefix served by a mobile router.

*Mobility Support Function (MSF)* – a set of processes executing at LER and responsible for mobile device (host or router) registration, Mobility Label assignment, Mobility Binding creation and distribution using a network update.

*Mobility Region* – a collection of the RAN cells or clusters served by a single MSF residing in the MPLS LER node. Each RAN cell or cluster is connected to the LER by means of the layer 2 grooming network (e.g. switched Ethernet) and terminates at the logical layer 3 interface of the LER that is under the control of MSF.

*Mobility Area* - a collection of Mobility Regions aggregated by the Area LER (ALER).

*Area-ID (AID)* – a unique identifier associated with a given Mobility Area. All MLBN nodes that belong to an Area are pre-configured with this identifier.

*Area LER (ALER)* – MPLS aggregation node that implements MSF and participates in the packet forwarding.

*Area Mobility Route Reflector (AMRR)* – a Mobility Route Reflector serving the Mobility Area. AMRR does not participate in the packet forwarding and performs control plane (signaling) functions using MP-BGP. All MSF LERs in the Mobility Area as well as the ALER(s) peer directly with the AMRR and act as the route reflector clients. The AMRR peers directly to all other AMRRs in the MLBN thus forming the mobility control plane hierarchy.

*MLBN Border Edge Router (MBER)* - Special types of MLBN node used to provide Inter-Carrier mobility management. MBERs establish mobility enabled MP-BGP peering points between the H-MLBN networks of different providers. These nodes perform a function similar to the ALER nodes and peer to their local AMRR nodes.

*Mobility Route Reflection* – a process of relaying Mobility Binding information to MLBN nodes without a requirement for full-mesh logical peering among all nodes. Route Reflector Clients (LER, ALER) peer directly to a Route Reflector (AMRR) but not with each other.

*Mobility Label (ML)* – a MPLS label that is associated with a mobile prefix (IPv4 or IPv6 MN's address or network prefix served by a mobile router). Mobility label is used to represent current network location of a mobile device. Mobility Labels are assigned by LERs and used as inner labels in the MPLS label stack.

*Local Mobility Label (LML)* – a ML that is locally significant at a given LER or ALER and is used to identify an intermediate MLBN node and a LSP segment leading toward a mobile prefix. LML is associated with a Current Mobility Label (CML).

*Current Mobility Label (CML)* – A ML that is associated with a mobile prefix and identifies RAN specific information to reach a mobile device or a next LSP segment toward the location of a mobile device. A CML may change as mobile devices move, while the associated LML may stay unchanged.

*Mobility Binding (MB)* – an association between the mobile prefix, MLBN LER Router-ID and the Mobility Label. Mobility Bindings are distributed using MP-BGP and encoded using a specific Network Layer Reachability Information (NLRI) format.

*Network Update* – a process of distributing Mobility Bindings throughout MLBN.

*Network Update Mode* – determines the destination of the network update in MLBN. Update Mode is encoded into the Mobility Binding message and is used to instruct a receiving MLBN node on how to forward the update to other nodes. Several Network Update Modes are used in MLBN:

- *Selective Downstream Push* – used to send Mobility Bindings to a given MLBN node (such as from LER to a local AMRR)
- *Unsolicited Downstream Push* – used to flood Mobility Bindings to a set of MLBN nodes (such as from an AMRR to all other AMRRs in MLBN, or from AMRR to all LERs in an Area)

*Network Update Type* – determines the scope of the Network Update in MLBN. Update Type is encoded into the Mobility Binding message and is used to instruct the receiving MLBN node on how to process the Mobility Binding. Several Network Update Types are used in MLBN:

- *Internal Update* – Used to distribute Mobility Bindings within a Mobility Area. Internal Update may be initiated by a LER and sent to a local AMRR to be reflected to a local ALER. Internal Update may also be initiated by an ALER in response to an External Update.
- *External Update* – Used to distribute Mobility Bindings outside a Mobility Area. External Update may be initiated by an ALER and sent to a local AMRR to be reflected to other AMRRs in MLBN or stored for subsequent Mobility Label Requests.
- *Inter-Carrier Update* – Used to distribute mobility bindings for mobile nodes that have roamed into the H-MLBN of another carrier. This update type instructs the AMRR or ALER to update an MBER node.

*Mobility Binding Request and Reply* – An explicit request for Mobility Binding information sent by a MLBN node. This request is carried using MP-BGP signaling and is formatted using specific NLRI encoding [2]. A Mobility Binding Reply is send by a MLBN node if the requested Mobility Binding for a mobile prefix in question exists.

*Last Requestor List (LRL)* - a list of Area IDs of the nodes that have requested Mobility Binding information for a particular mobile node during the lifetime of the corresponding Mobility Binding. LRL is maintained by an AMRR. Two types of LRLs are defined in MLBN:

- *External LRL (eLRL)* - is a LRL that consists of the Area IDs of external AMRR nodes.
- *Internal LRL (iLRL)* - is a list of the Router IDs of all LER nodes internal to an area which originated Mobility Binding requests for mobile prefix in question.

*Segmented LSP* – a sequence of MPLS LSPs used to reach a LER node serving a mobile device and a corresponding mobile prefix. Selection of a next LSP segment is based on the Current Mobility Label associated with a mobile prefix at an intermediate H-MLBN node. The switching to the next LSP segment at an intermediate H-MLBN node is performed using MPLS operations (pop, swap, push) and without the IP address lookup.

Fig. 5 illustrates a Hierarchical MLBN (H-MLBN) with nine Mobility Regions (MR) served by the MSF LER nodes at the edge level of the MPLS network. Each LER connects to a set of RAN cells or clusters using the layer 2 access/grooming network such as Ethernet or ATM (not shown). Each RAN cell or cell cluster is terminated at a layer 3 logical interface of the LER controlled by the MSF as shown in Fig. 2 (Mobility Region (n-1)).

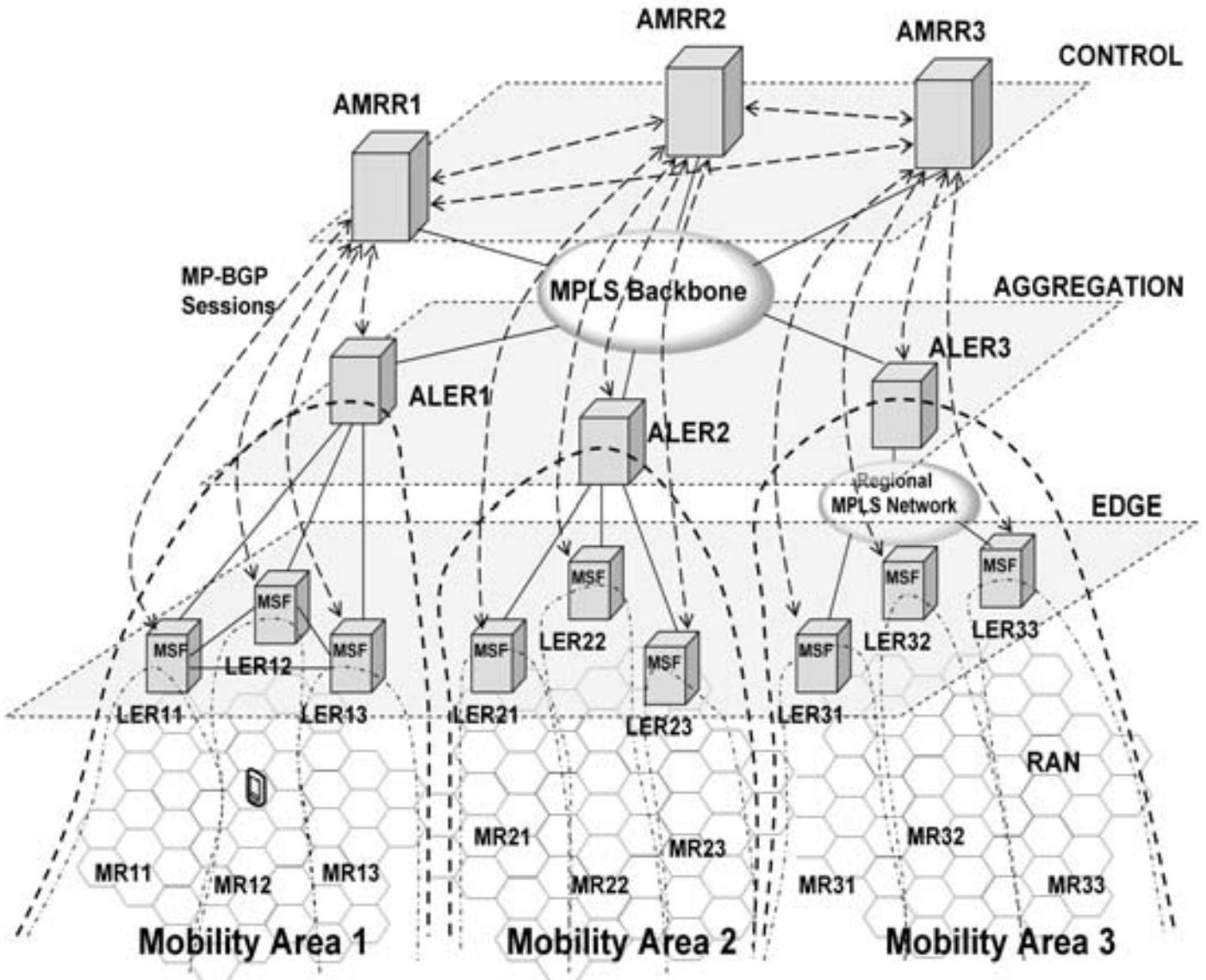

Figure 5. Hierarchical Mobility Label Based Network.

At the aggregation level the edge level LER nodes are connected to the Area LER nodes (ALER) thus forming the Mobility Areas. Fig. 5 shows three Mobility Areas each comprising of three Mobility Regions. Please note that the edge LER nodes do not have to be directly connected to the ALER node. A regional MPLS network may be used to provide this connectivity (as shown in Mobility Area 3 in Fig. 5).

At the control plane level Fig. 5 shows three Area Mobility Route Reflectors (AMRR) serving their respective Mobility Areas and peering directly with the corresponding MSF LER and the ALER nodes within the area. The MSF LERs and the ALER act as route reflector clients. The AMRR nodes also peer directly with all other AMRR nodes forming a logical full mesh of the MP-BGP control plane sessions. The ALER and AMRR nodes are connected to the MPLS backbone network that consists of the LSR nodes.

The mobile hosts and routers register only with the serving MSFs at the edge level of the H-MLBN. The MSF Discovery and Registration protocols defined in [2] do not extend beyond the MSF LER nodes. The registration with a MSF results in the assignment of the Mobility Label and generation of the corresponding Mobility Binding as described in [2].

Upon the completion of the registration the LER nodes at the edge level update the AMRR using a Selective MP-BGP update mode [1-2] carrying the Mobility Binding information for the registered mobile devices. In contrast to the regular BGP Route Reflector operation, the AMRR does not automatically update all of its peers. The automatic or unsolicited reflection of the Mobility Binding information is only executed toward the ALER node within the Mobility Area. A given AMRR node does not automatically update its LER clients or the rest of its peer AMRR nodes, they rely on the On-Demand Request/Response mechanism in order to acquire the needed Mobility Bindings.

The Mobility Binding updates may be internal or external. An internal update is initiated by an LER node local to an area and carries the Mobility Binding information for a locally registered mobile device. The internal update is sent by an LER to the AMRR in order to update the ALER node. The internal update may also be sent by the ALER node to AMRR in response to the external update received by the ALER about the Mobility Bindings originating outside a local area. An external update is originated by the ALER in response to an internal update and is sent to the AMRR. The combination of internal and external updates allows to maintain local significance of Mobility Labels as well as the implementation of the hierarchical packet forwarding as will be shown in the examples below. The update types are shown in Fig. 6.

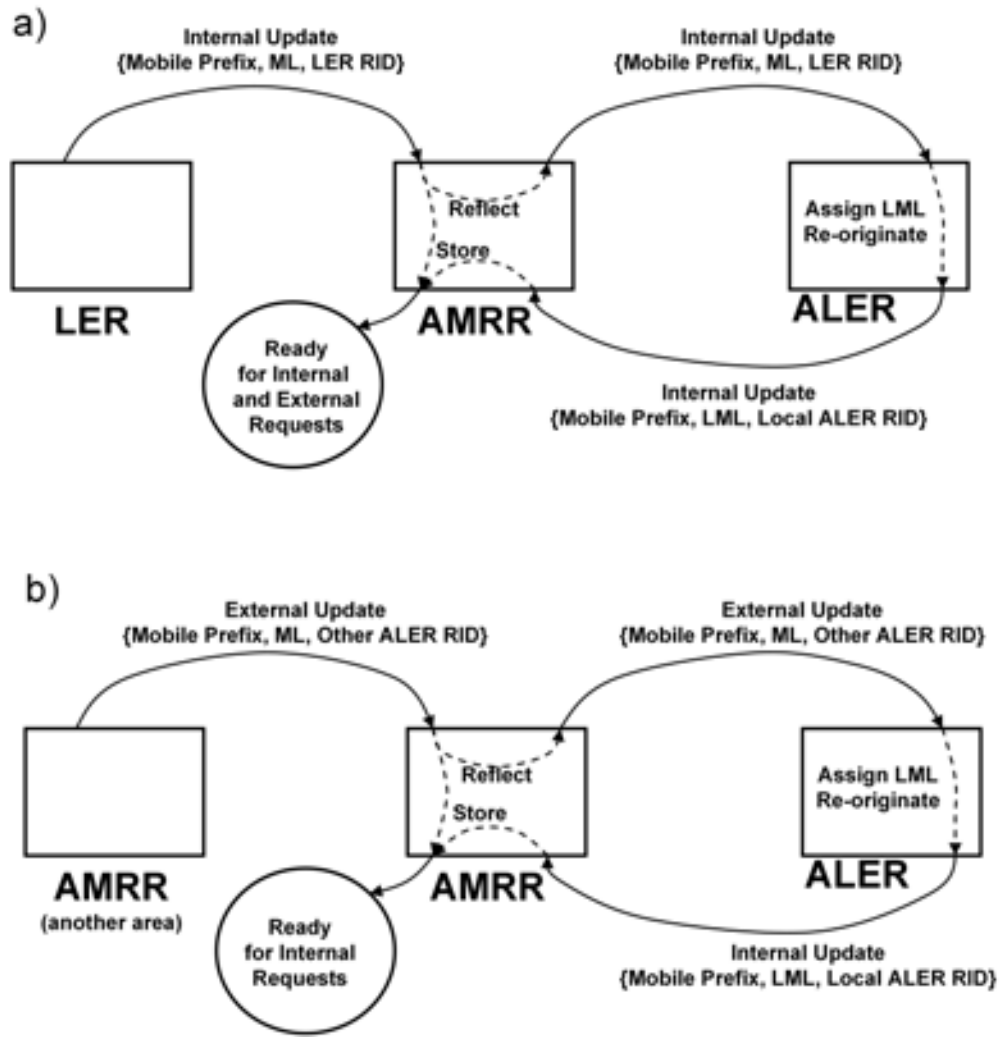

Figure 6. MLBN Update Types. Internal update processing (a), external update processing (b). ML – Mobility Label, LML – Local Mobility Label, RID – Router ID.

In addition the following architectural differentiation allows implementation of a forwarding plane hierarchy that enables additional scalability efficiencies. The AMRR node may explicitly distinguish between the two types of the route reflector clients – the set of the LER nodes and the set of the ALER nodes. Each client is identified to the AMRR by its Router ID. When a LER updates AMRR using the internal update for a locally registered mobile, AMRR uses the originating LER's Router ID as the "Origin MP-BGP Next-Hop" in the Mobility Binding to update other area LER nodes and the ALER. When ALER receives the internal update it stores the Mobility Binding, assigns a new and unique Local Mobility Label and re-originates the Mobility Binding using the external update to send the information to AMRR using its own Router ID (as opposed to the originating LER's Router ID) as the "Origin MP-BGP Next-Hop".

When AMRR receives a Mobility Binding from another peer AMRR node outside the local area, AMRR sends an external update to ALER (this update carries the Mobility Label assigned by the outside area ALER and uses that ALER's Router ID as the "Origin MP-BGP Next-Hop". After receiving the external update, ALER stores the Mobility Binding, assigns a new and unique Local Mobility Label and re-originates the Mobility Binding using the internal update to send the information to AMRR using its own Router ID. AMRR uses this information to update the local LER nodes.

In other words, all Mobility Bindings for the mobile nodes residing in a given Mobility Area are represented to other areas as originated by the local ALER node, and all Mobility Bindings for the mobile nodes residing outside the local area are represented as reachable via the local ALER node.

From the forwarding plane perspective this means that the MPLS packets carrying a label stack in their headers and sourced by the LER nodes in a given Mobility Area will be directed to their local ALER. The local ALER node will pop the top MPLS label (the outer label) as it is always the terminating node for the MPLS LSP and will examine the next label in the stack – the Mobility Label to identify the far end ALER node in the destination Area. The local ALER will impose the label stack and forward the packet to the far end ALER. The far end ALER will also pop the top label (as it is the terminating node for the LSP), read the Mobility Label and identify the individual LER node and the corresponding MSF serving the current location of the mobile node.

It is important to point out that the label operations (pop, swap and push, both for the infrastructure and mobility labels) at ALER do not involve a lookup of an IP prefix for a given mobile device. These operations are driven by a Forwarding Information Base (FIB) structure maintained at ALER and managed by the control plane.

If the traffic between the mobile devices is contained within a given Mobility Area and has to traverse the ALER node, no Mobility Label inspection by the ALER node is necessary (as the ALER will not be terminating the LSP between two LER nodes within the same Mobility Area).

As shown in Fig. 7, traffic between $MN_2$ and $MN_3$ uses a Pass-Trough LSP via ALER (switched using top labels). Traffic between $MN_4$ and $MN_1$ uses a segmented LSP: local ALER terminates the LSP, looks up Mobility Label and forwards to the far end ALER. The far end ALER terminates the LSP, looks up Mobility Labels and forwards traffic on a new LSP toward $MN_1$.

It is assumed that there exists any-to-any MPLS LSP connectivity among the Router IDs of all of the LER nodes of the network (including the ALER nodes) using the infrastructure (outer or top) MPLS labels. It is also assumed that the mobile nodes move randomly between the cells of the RAN within a given Mobility Region. The transfers from one Mobility Region to another and from one Mobility Area to another occur in a sequential manner. For example, in Fig. 5, a mobile device in MR12 may transfer to MR11 or MR13 but not directly to MR21 (this type of transfer is considered to be a "start" condition – meaning that it is treated as an arrival of a new mobile device).

Clearly the strategy to update the rest of the network may have numerous variations. The update strategy described below is based on the combination of the On-Demand Request/Response transactions as well as the Unsolicited Downstream and Selective (targeted) updates. In addition, to facilitate the hand-off process in the hierarchical environment the Last Requestor List (LRL) is introduced and associated with each Mobility Binding at the AMRR level. The LRL is a list of Area IDs of the AMRR nodes that have requested Mobility Binding information for a particular mobile node during the lifetime of the corresponding Mobility Binding.

In what follows we describe various hand-off scenarios and the corresponding operation at the control and forwarding plane levels. In all described cases we assume that the correspondent node (CN) is stationary and is located in MR33 (see Fig. 5).

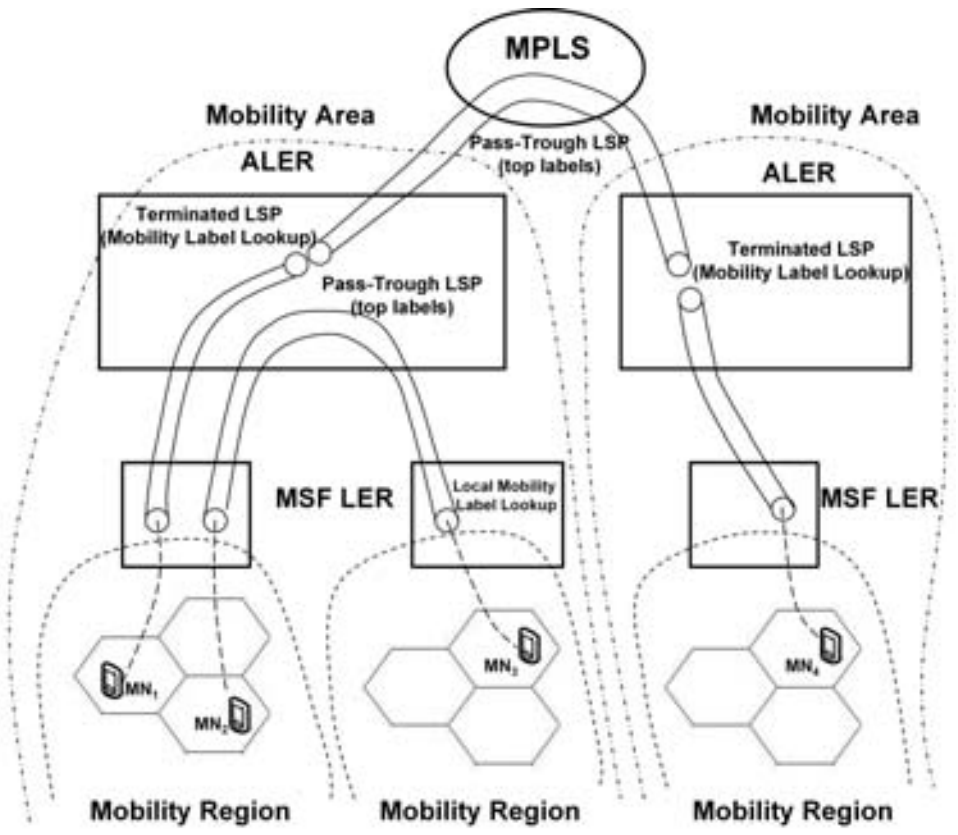

Figure 7. Hierarchical Mobility Forwarding Plane.

### A. Start-Up

The sequence of events that occurs when a mobile device is first turned on or when a mobile device re-initializes is referred to as a start-up. Consider the H-MLBN shown in Fig. 5, where a mobile device is turned on in one of the RAN cells in MR12.

1. MN initiates the MSF Discovery and receives the Virtual Link Layer and IP Addresses of the serving MSF located in LER12.
2. MN initiates the MSF Registration and communicates its IP address to the MSF. The registration messaging also includes the Area ID information. In a start condition the mobile device must use the Area ID value of 0. In response the MSF communicates to the mobile a new Area ID value of 1 (LER12 serves MR12 in Area 1) and other related information (see [2]).

3. LER12 creates a Mobility Binding for the MN and updates AMRR1 using the internal update. The Mobility Binding carries the MN's IP address, the LER12's Router ID, the Mobility Label and the Area ID value of 1.
4. AMRR1 stores the received binding information for the MN (including the received Area ID value from LER) and associates an empty LRL with the Mobility Binding.
5. AMRR1 updates ALER1 with the MN's Mobility Binding using the internal update. The associated LRL is not sent to ALER1 and is stored locally at AMRR1. ALER1 receives the update and allocates a Local Mobility Label. This label must be unique within the ALER. ALER1 updates its Forwarding Information Base (FIB) and creates a label trail. The trail record consists of the incoming and outgoing Infrastructure (outer or top) labels and the Local and Current Mobility Labels (inner) labels. The incoming and outgoing outer labels are associated with the "Origin MP-BGP Next-Hop" received in the Mobility Binding for the MN and are exactly the same labels used to reach the Router ID of LER12 distributed by the Label Distribution Protocol (LDP). The Current Mobility Label is the one received in the Mobility Binding update from AMRR1. The Local Mobility Label is used to represent the MN to the outside areas. The sample FIB structure is shown in Fig. 8.

| | |
|---|---|
| Mobile Prefix (FEC) | 10.1.1.1/32 |
| Origin Router ID | 20.1.1.12 |
| In Top Label | 16 |
| Local Mobility Label | 216 |
| Current Mobility Label | 116 |
| Out Top Label | 17 |
| Out Interface ID | GIG1/0/3 |

Figure 8. Sample ALER Forwarding Information Base structure.

6. ALER1 updates AMRR1 using external update with the Mobility Binding for MN. The binding carries the MN's IP address, the Local Mobility Label assigned by ALER1 and the ALER1's Router ID as the "Origin MP-BGP Next-Hop". AMRR1 does not need to store a separate external Mobility Binding for the MN. It adds the Local Mobility Label to the record and uses it and the ALER1 Router ID in replies to the requests from outside the area.
7. Assume that a CN located in MR33 sends a packet to the MN in MR12. The packet reaches LER33.
8. LER33 identifies that the destination IP address in the packet belongs to the mobility address range, looks up its existing Mobility Binding and finds no matches. LER33 requests the Mobility Binding information for the MN from its AMRR3 using the On-Demand Mobility Binding Request [2]). LER33 uses the Area ID value of 3 in this request.
9. Since AMRR3 does not have the Mobility Binding for the MN it forwards the request to both AMRR2 and AMRR1. AMRR1 replies with the Mobility Binding. *AMRR1 uses the ALER1 Router ID (as opposed to the LER12 Router ID) as the value of the "Origin MP-BGP Next-Hop" and the Local Mobility Label assigned by ALER1 in the Mobility Binding sent to AMRR3. The Area ID value of 1 is also included into the binding information.* AMRR1 populates the LRL with the Area ID value of 3. *In order to avoid traffic looping conditions, the AMRR nodes send positive replies to the binding requests only for the Mobility Bindings that have their Area ID matching the replying AMRR node's Area ID.*
10. AMRR3 reflects the reply from AMRR1 to ALER3 with the MN's Mobility Binding information using external update. ALER3 receives the update and allocates a Local Mobility Label. This label must be unique within the ALER. ALER3 updates its Mobility Forwarding Information Base (FIB) using the received Mobility Binding. In other words ALER3 creates the Mobility Label trail described in step 5 and shown in Fig. 8.
11. ALER3 updates AMRR3 with the MN's Mobility Binding using internal update. The binding carries the MN's IP address, the Local Mobility Label assigned by ALER3 and the ALER3's Router ID as the "Origin MP-BGP Next-Hop".
12. AMRR3 updates LER33 with the Mobility Binding received in an internal update from ALER3.
13. LER33 imposes the label stack on the received IP packet from the CN. The outer label is the one associated with the "Origin MP-BGP Next-Hop" listed in the Mobility Binding for the MN (*this is the same as the Router ID of ALER3*). The inner label is taken directly from the Mobility Binding. LER33 sends the MPLS frame downstream to ALER3.
14. ALER3 pops the outer (top) label (as it is the terminating node for the LSP) and looks up the Mobility Label. ALER3 locates the FIB record containing the value of the Local Mobility Label found in the packet, and uses the value of the corresponding Current Mobility Label to swap Mobility Labels (in this case it is equal to the value of the Local Mobility Label assigned to the MN by ALER1). ALER3 also uses the top label associated with the Current Mobility Label and pushes it onto the packet's label stack. ALER3 forwards the packet onto the LSP identified by the new top label that is based on the Forwarding Equivalency Class (FEC) associated with the Router ID of ALER1.

15. The packet reaches ALER1. ALER1 reads the top label and establishes that it is the terminating point for this LSP. ALER1 pops the top label and proceeds to read the next label in the stack – the Mobility Label. ALER1 looks up the FIB based on the value of the Local Mobility Label. Once the value of the Local Mobility Label is located, ALER1 finds the corresponding values of the Current Mobility Label and the Out Top Label associated with it. In this case the value of the Current Mobility Label corresponds to the value of the Mobility Label assigned to the MN by LER12, and the value of the Out Top Label is associated with the FEC for the Router ID of LER12. ALER1 imposes the label stack on the outgoing packet and forwards the packet on to the LSP toward LER12.
16. LER12 pops the outer label (in some cases this label may be replaced by an implicit null label by ALER1 as it may be the Penultimate Hop for this LSP). LER12 reads the Mobility Label, looks up its MSF database and locates the record associated with the mobile node. This record may include all the necessary layer 2 information specific to the RAN in which the mobile is located. The packet is then forwarded out the logical layer 3 interface associated with the mobile node.

The Start-Up process is shown in Fig. 9 below.

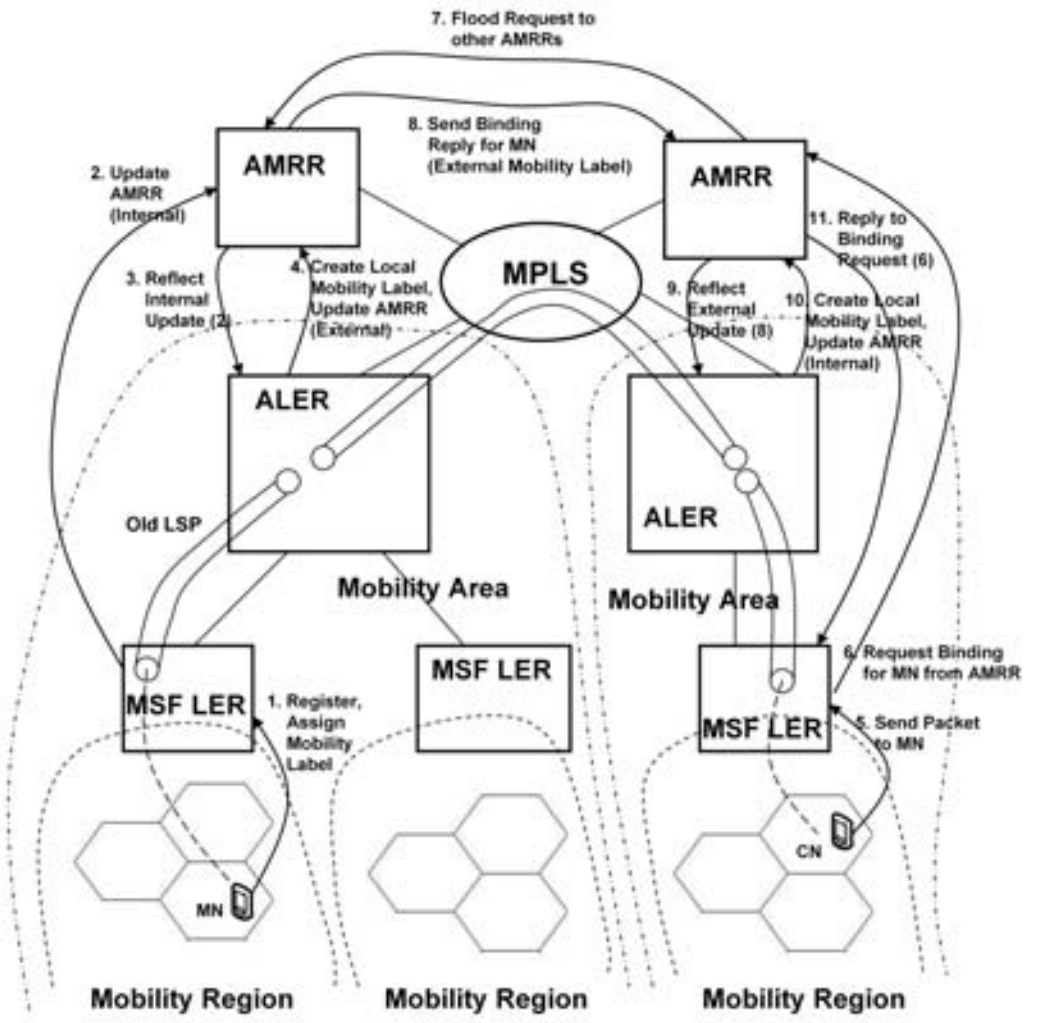

Figure 9. Illustration of the Start-Up sequence.

The timing diagram of the Start Up control sequence and the associated forwarding steps is shown in Fig. 10.

*B. MSF-Local hand-off*

This type of hand-off corresponds to the mobile device movements contained within a given Mobility Region, where the mobile is moving among the RAN cells or clusters served by a single MSF. The following sequence of signaling steps is performed:

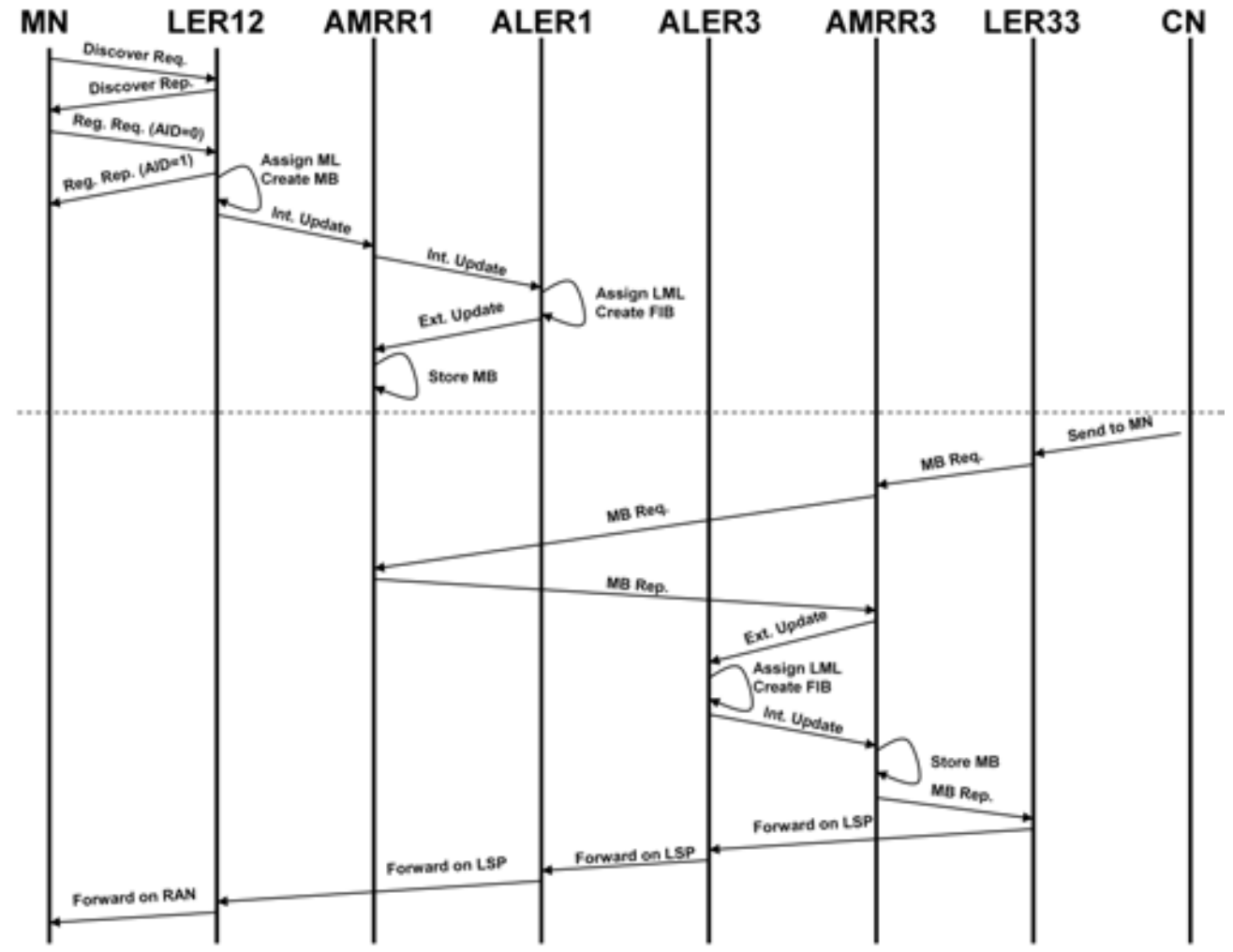

Figure. 10. Timing diagram for the Start Up sequence in H-MLBN

17. Assume that the MN now moves into another RAN cell or cell cluster in the same MR12 (Fig. 5). After the radio hand-off is completed the MN should either continue sending packets to the CN (using the virtual link layer address of the MSF) or if there are no packets to send it should be sending periodic registration keepalives to the serving MSF's virtual link layer address (the keepalive messages carry the Area ID value of 1 communicated to mobile in step 2).
18. The MSF in LER12 "tracks" the mobile node by making note that the packets with the mobile node's source link layer and IP addresses started arriving on a different layer 3 logical interface (associated with the new RAN cell or cluster). The MSF then updates the local association table in the LER12 with the new layer 3 interface ID for the mobile node. Alternatively, the association record may be updated based on the reception of the keepalive messaging from the mobile node (since this messaging carries the Area ID value of 1 the MSF logic may be able to determine and log the MSF-Local hand-off event for the mobile node for reporting purposes). If the mobile node moves into the RAN with different layer 2 characteristics from the original RAN, the MSF may update the local association record with the layer 2 specific information (such as encapsulation and the link layer headers). This is illustrated in Fig. 11.
19. Traffic delivery to the new location of the mobile node follows the same process as described in steps 13 – 16. Note that the Mobility Label for the mobile node did not change and that no Mobility Binding updates were necessary.

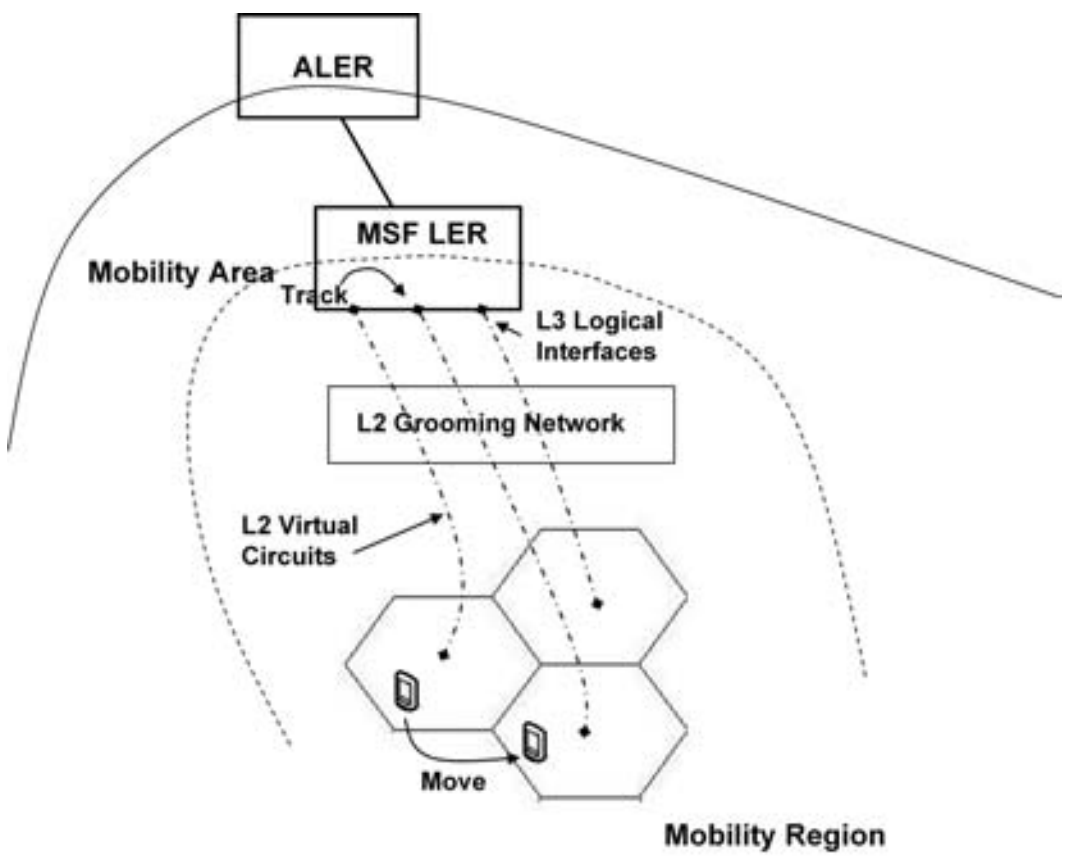

Figure 11. MSF-Local Hand-Off.

*C. Inter-MSF Intra-Area hand-off*

This hand-off takes place when a mobile node moves from one Mobility Region to another but stays within a given Mobility Area. In other words the mobile transfers from the control of the MSF in one LER to the control of the MSF of another LER with both LERs connected to the same ALER. Consider the H-MLBN shown in Fig. 5, where a mobile device continues to move from a RAN cell in MR12 to a RAN cell in MR13. We continue with the example where a CN in MR33 is communicating with the MN.

When MN transfers to a different MR, initially it has no indication that it is under the control of a new MSF and it continues to either send packets to its destination or keepalives to the virtual link layer address of the MSF. Generally the RAN technology in the new MR may be different than the one in the old MR and in this case the link layer addressing may also be different. Therefore, generally it may be assumed that the mobile will need to execute a new Discovery and Registration procedure with the new MSF. However, in cases where the same RAN technology and the same link layer addressing is used in the grooming network, a virtual addressing scheme may be implemented in which all mobiles use the same MSF virtual link address throughout the network (this is similar to the use of virtual addresses in the router redundancy protocols such as VRRP or HSRP). Assuming that the new registration is required, and continuing with the example, the following sequence of events takes place:

20. MN initiates the registration with the new MSF in LER13 and communicates the last Area ID value of 1 in the process. LER13 also uses the same value in the registration and keepalive messaging with the MN.
21. LER13 updates AMRR1 with the Mobility Binding for the MN including the new Mobility Label and the Area ID value of 1 received from the MN using internal update. LER13 assigns a locally significant Mobility Label to the MN that is now registered to its MSF.
22. AMRR1 reads the Area ID value and compares it with the last recorded Area ID for the MN. If the values are the same, AMRR1 reflects the Mobility Binding to ALER1 without looking up the LRL and any further update action towards other peer AMRR nodes. AMRR1 may also reflect the binding to LER12 since the “Origin MP-BGP Next-Hop” changed in the Mobility Binding from LER12 to LER13 Router ID.
23. ALER1 receives the internal update and in turn updates the label trail record for the MN in its FIB. As shown in Fig. 8 the values of Router ID, In Top Label, Current Mobility Label, Out Top Label and the corresponding Interface ID values are updated. ALER1 does not execute the external update to AMRR1 since the Mobility Binding for the MN already exists and the Local Mobility Label has already been assigned.
24. CN in MR33 continues to send packets to the IP address of the MN. Both LER33 and ALER3 use their existing label stacks to forward the packets onto the LSP leading to ALER1.
25. ALER1 receives the MPLS packet and pops the top label (this is the incoming top label associated with its own Router ID). Then ALER1 reads the Mobility Label and looks up its FIB to find the trail record with the matching Local Mobility Label value. Once the record is located ALER1 swaps the packet’s Mobility Label with the value of the Current Mobility Label and pushes the associated Out Top Label onto the label stack. The top label will be the one associated with the Router ID of LER13 based on the “Origin MP-BGP Next-Hop” in the received Mobility Binding (step 23).
26. LER13 receives the frame and repeats the process described in step 16 to deliver the packet to the MN.

The Inter-MSF Intra-Area hand-off is illustrated in Fig. 12. Note that this hand-off does not require a network update outside of a given Mobility Area.

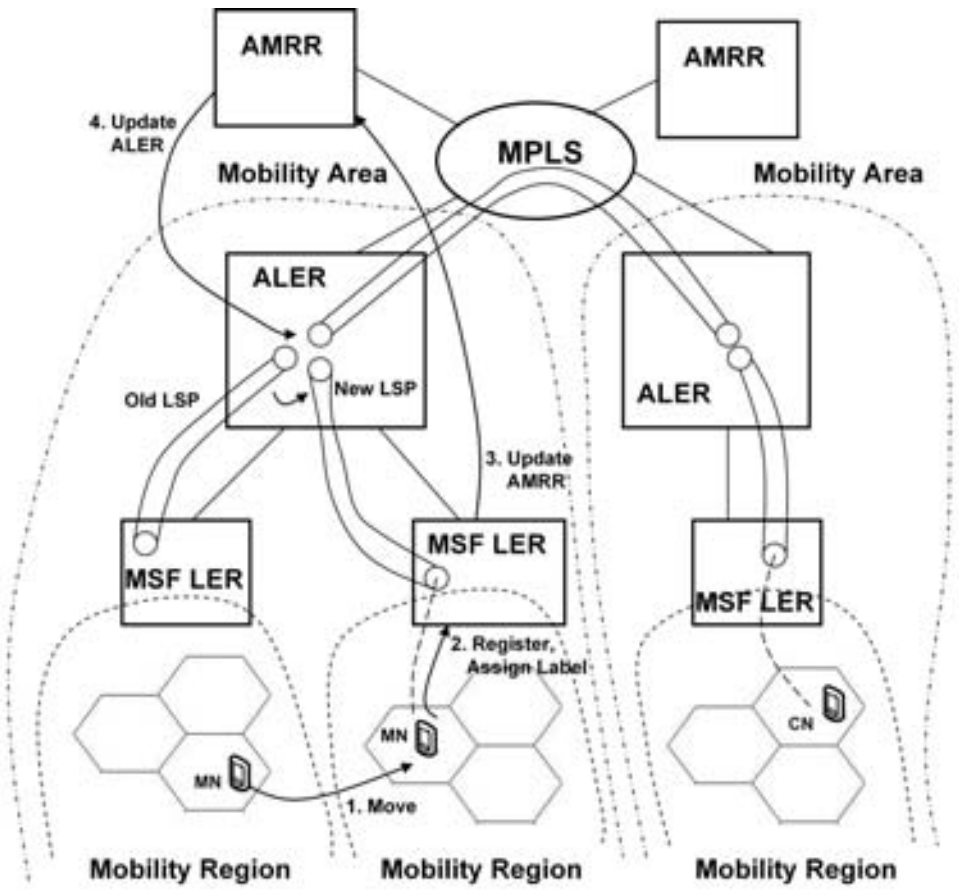

Figure 12. Inter-MSF Intra-Area Hand-Off.

The timing diagram of the Inter-MSF Intra-Area hand-off control sequence and the associated forwarding steps is shown in Fig. 13.

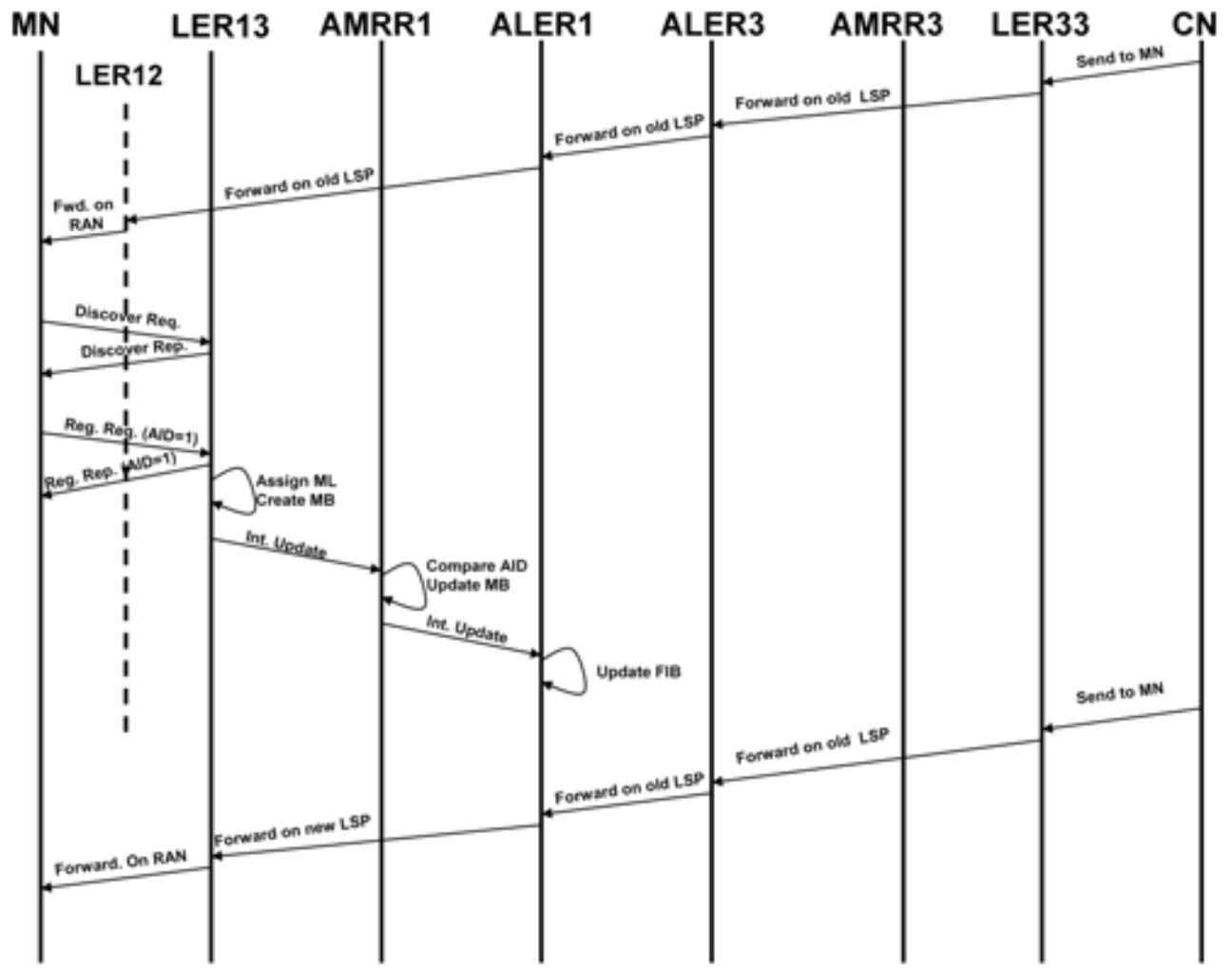

Figure. 13. Timing diagram for the Inter-MSF Intra-Area hand-off sequence in H-MLBN

*D. Inter-MSF Inter-Area hand-off*

The Inter-Area hand-off occurs when a mobile device moves to a new Mobility Region that is also part of a new Mobility Area. Referring to Fig. 5, if a mobile node moves from a RAN cell in MR13 to a RAN cell in MR21 it transitions from the control of the MSF in LER13 to the control of the MSF in LER21. Since these two LER nodes are connected to two different ALER nodes this type of move is referred to as the Inter-Area hand-off.

Continuing the previous discussion we describe the sequence of events when the MN keeps on moving and transitions from MR13 to MR21 and maintains communication with the CN in MR33. As earlier, we assume that the MN initiates a new Discovery and Registration process in MR21.

24. MN initiates the registration with the new MSF in LER21 and communicates the last Area ID value of 1. LER21 sends the Area ID value of 2 to the MN in the registration and keepalive messaging.
25. LER21 updates AMRR2 using internal update with the Mobility Binding for the MN including the new Mobility Label and the Area ID value of 1 received from the MN.
26. AMRR2 reads the Area ID value and compares it with the last recorded Area ID for the MN. Since there may not be a record for the MN, AMRR2 also compares the received Area ID (1) with its own Area ID (2). Since the values are different, AMRR2 determines that it needs to send a LRL request to AMRR1. AMRR2 updates the Area ID value for the MN to its own Area ID (2).
27. AMRR2 updates ALER2 with the Mobility Binding for the MN using internal update. ALER2 assigns a Local Mobility Label and creates the label trail record in the FIB (Fig. 8).
28. ALER2 updates AMRR2 with the Mobility Binding for the MN carrying the MN's IP address, the Local Mobility Label and ALER2's Router ID as the "Origin MP-BGP Next-Hop' using external update.
29. AMRR2 sends the LRL Request [2] to AMRR1 (last recorded AMRR for the MN). In addition, AMRR2 sends a Mobility Binding Update to AMRR1 along with the LRL Request. The update carries the MN's IP address the Local Mobility Label assigned by ALER2, ALER2's Router ID as the "Origin MP-BGP Next-Hop" and the Area ID value of 2.
30. AMRR1 replies to AMRR2 with the LRL for the MN. The LRL contains the last requestor information: the Area ID value of 3. AMRR2 updates the local LRL with the received information.
31. AMRR1 updates ALER1 with the received binding update from AMRR2 using external update. ALER1 updates the Current Mobility Label, the Router ID and the corresponding top labels in the FIB record for the MN. ALER1 does not re-originate the internal update since the Local Mobility Label already exists for the MN.
32. AMRR2 receives the LRL Reply from AMRR1 and sends the Mobility Binding update for the MN to AMRR3 using the ALER2 Router ID as the value for the "Origin MP-BGP Next-Hop" and the Local Mobility Label assigned by ALER2.
33. AMRR3 receives the update and reflects the Mobility Binding to ALER3 without updating any other nodes in the Area.
34. ALER3 updates its label trail with the new value of the Current Mobility Label taken from the Mobility Binding update from AMRR3. ALER3 also updates the values of the Router ID (ALER2) and the associated top labels.
35. LER33 uses the existing label stack when encapsulating the packets from the CN to the MN.
36. ALER3 receives the packets and pops the top label (since it terminates the LSP), looks up the FIB to locate the Local Mobility Label and swaps the Mobility Label in the packet with the value of the Current Mobility Label found in the record. ALER3 also locates the corresponding Out Top Label, imposes the label stack on the packet, and forwards it onto the LSP toward ALER2.
37. Traffic delivery follows the steps 25 – 26 using ALER2 instead of ALER1 and LER21 instead of LER13.

Inter-Area hand-off is illustrated in Fig. 14. It is important to note that during the transient conditions there exist a possibility for communications between MN and CN with minimal disruptions. For example, during the initial layer 2 hand-off the MN may communicate to two RAN base stations at the same time (soft hand-off). In this case, while MN performs the Discovery and Registration with the new MSF the packets from CN may still be delivered using the old segmented LSP. In addition, once ALER1 has been updated (step 31) and before ALER3 is updated, traffic from CN to MN that is sent on to the LSP from ALER3 to ALER1 may be re-routed by ALER1 to ALER2 using the new Current Mobility Label (shown in Fig. 15). Although this re-routed LSP represents a sub-optimal routing path, this condition is temporary – until the moment ALER3 is updated in step 34.

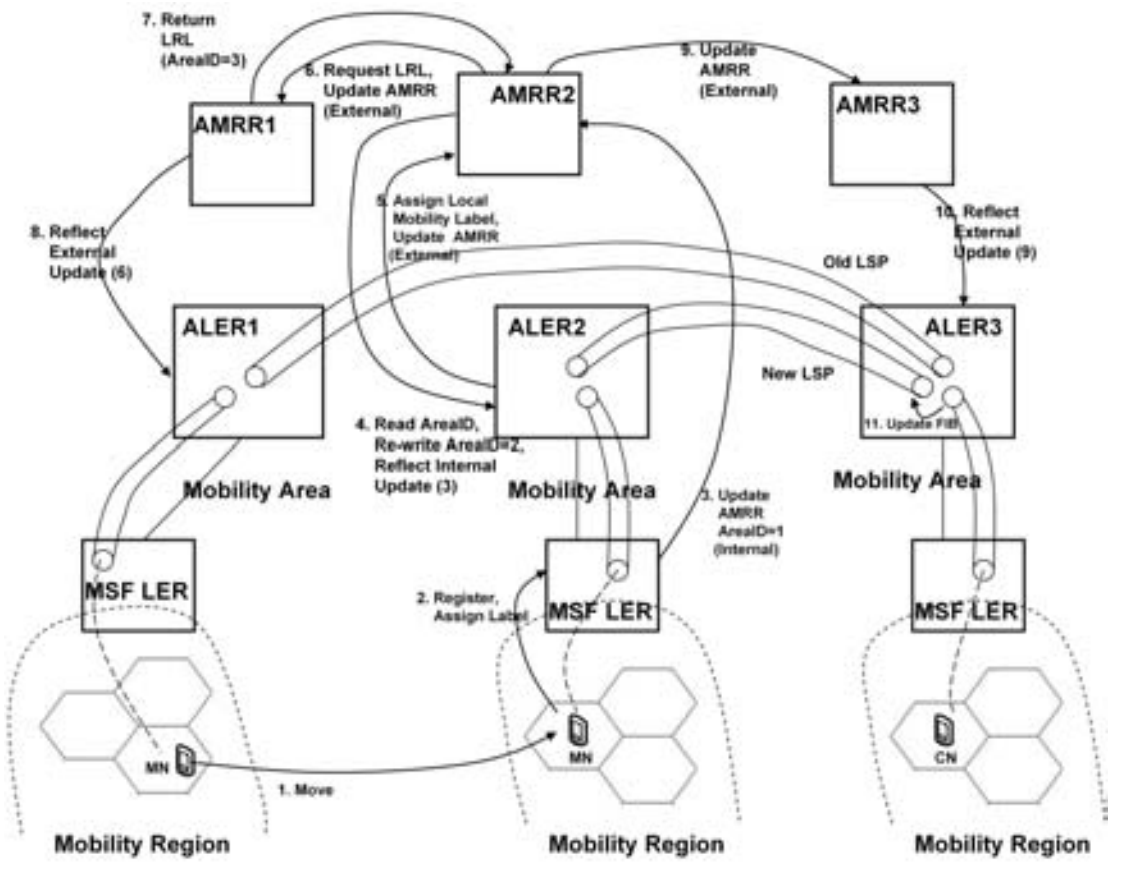

Figure 14. Inter-MSF Inter-Area Hand-off.

The timing diagram of the Inter-MSF Inter-Area hand-off control sequence and the associated forwarding steps is shown in Fig. 15.

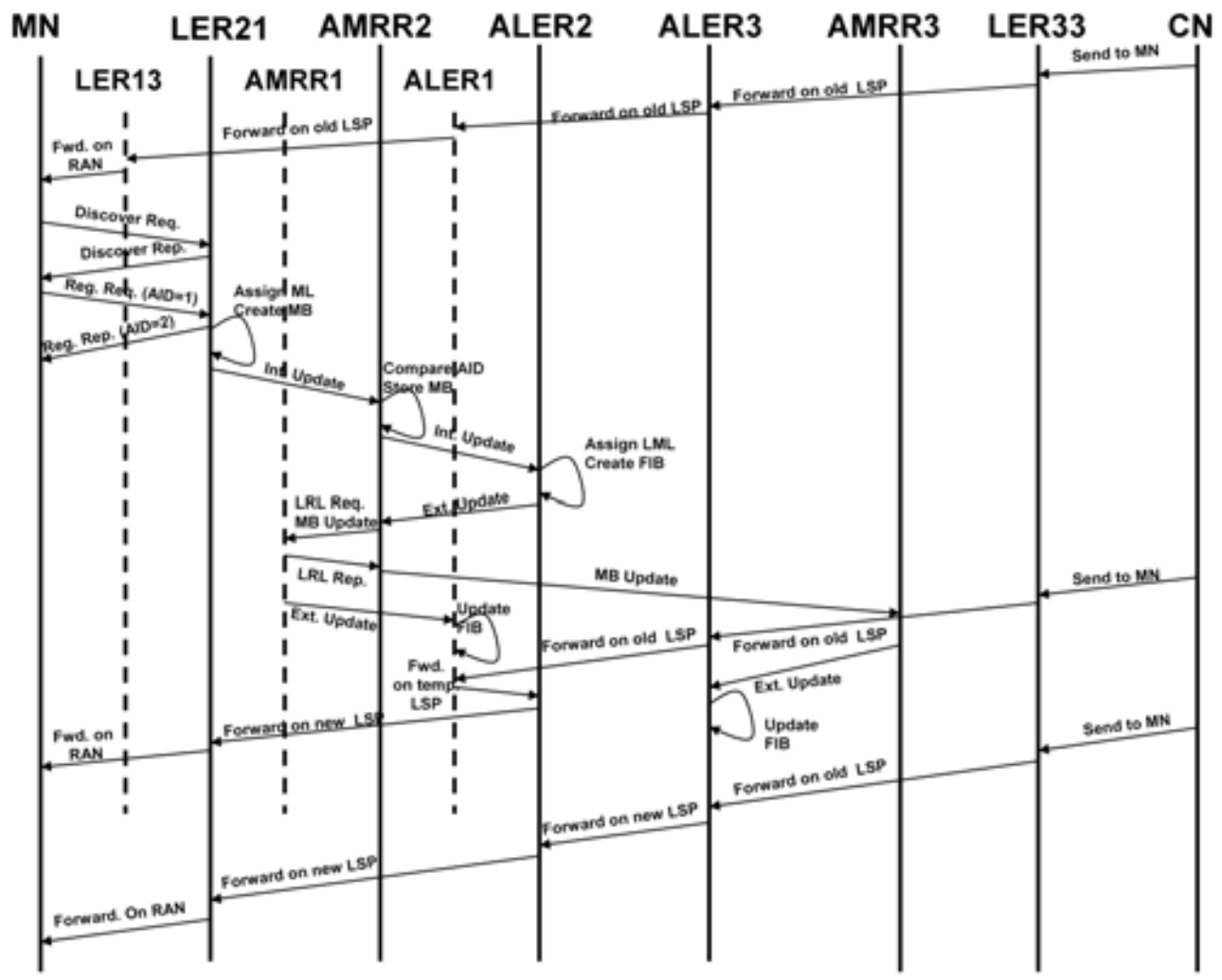

Figure. 15. Timing diagram for the Inter-MSF Inter-Area hand-off sequence in H-MLBN

## V. Discussion

### *A. Need for AMRR*

The main function of AMRR nodes is to act as a centralized control plane entity within a Mobility Area. This involves reflecting of internal updates from the LER nodes to the ALER node, processing of external updates from the local ALER node, reflection of external updates from outside the area to the ALER node, processing of internal updates from the ALER node and generation of Mobility Binding and LRL requests and replies. A separate AMRR node does not participate in packet forwarding and offloads these functions from the ALER nodes. This allows distribution of processing load and increased scalability. Fig. 16 shows the state diagram of the interactions among the LER, ALER and AMRR nodes.

At the same time the presence of AMRR requires additional signaling steps during the update process and increases the complexity of the overall scheme.

It is important to note that the functions of the AMRR node may be combined with the functions of the ALER node. This will avoid additional update steps following the allocation of the Local Mobility Label or reception of the Mobility Binding updates from outside the area. Clearly this will also increase the processing power requirements for the ALER nodes. In addition, implementation of certain architectural requirements such as survivability and load distribution may be complicated without a separate control plane layer represented by the AMRR nodes.

### *B. Survivability and Load Distribution*

In analyzing the H-MLBN architecture, natural questions to ask are: "what happens if an AMRR fails?" and "what happens if an ALER fails?". The ready answer is to use pairs of AMRRs and ALERs in the solution. However this also brings up the information synchronization and processing load control issues. The synchronization of information requires updating of both of the AMRR and ALER nodes in the high availability pair so that the fail-over may occur with the preservation of the state information on the existing Mobility Bindings. The load control has to do with the obvious desire to utilize the resources of both systems involved in the high availability configuration. An example of high availability structure is shown in Fig. 17.

In Fig. 17, each LER serving a Mobility Region is peering with both AMRR nodes serving the Mobility Area: Primary AMRR and Secondary AMRR. In addition, both ALER nodes: Primary ALER and Secondary ALER peer with both AMRR nodes, and the AMRR nodes peer with each other and the AMRR nodes in other Mobility Areas.

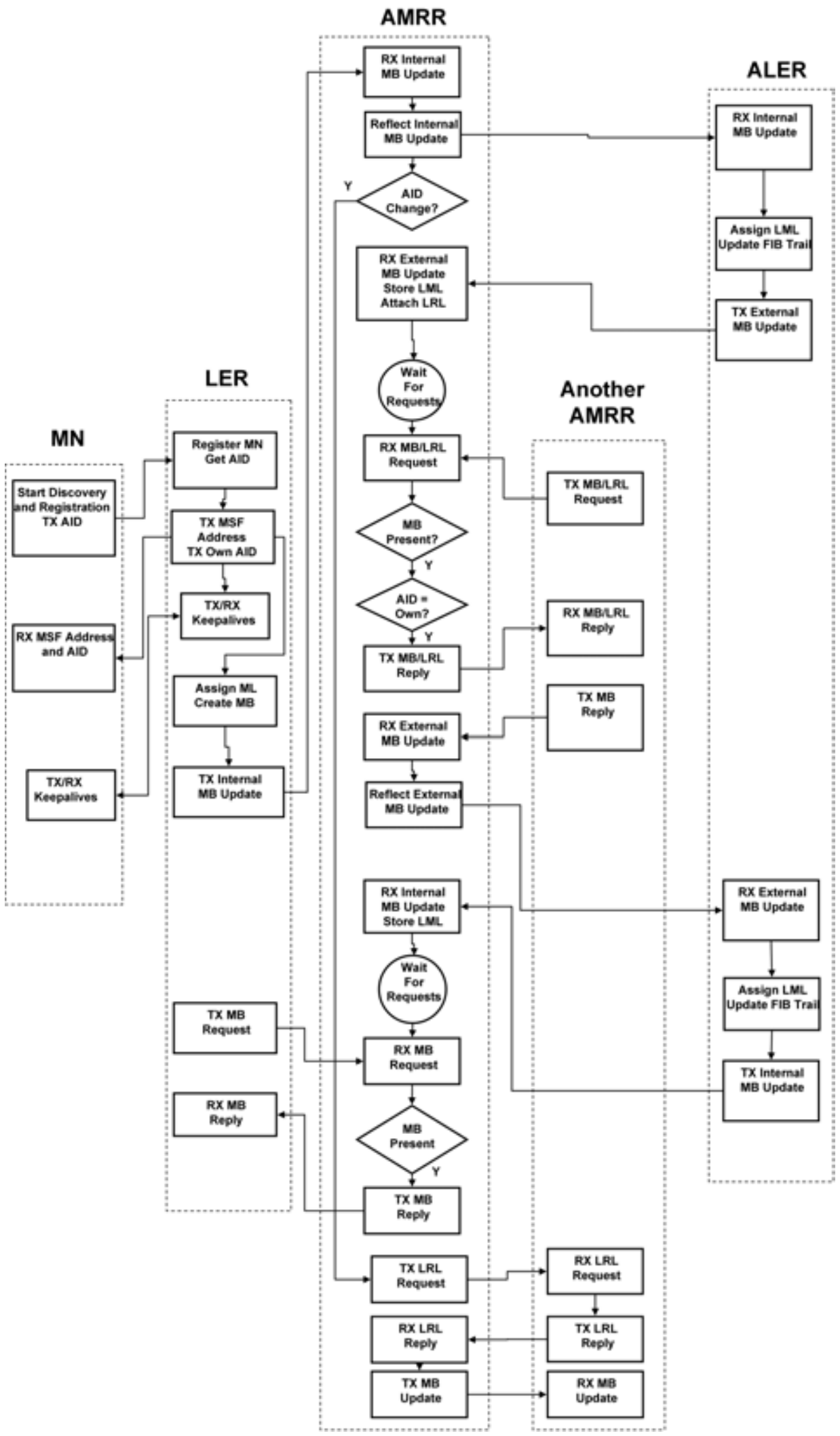

Figure 16. LER, ALER and AMRR interaction state diagram. AID - Area ID, MB – Mobility Binding, ML – Mobility Label, LML – Local Mobility Label, LRL – Last Requestor List.

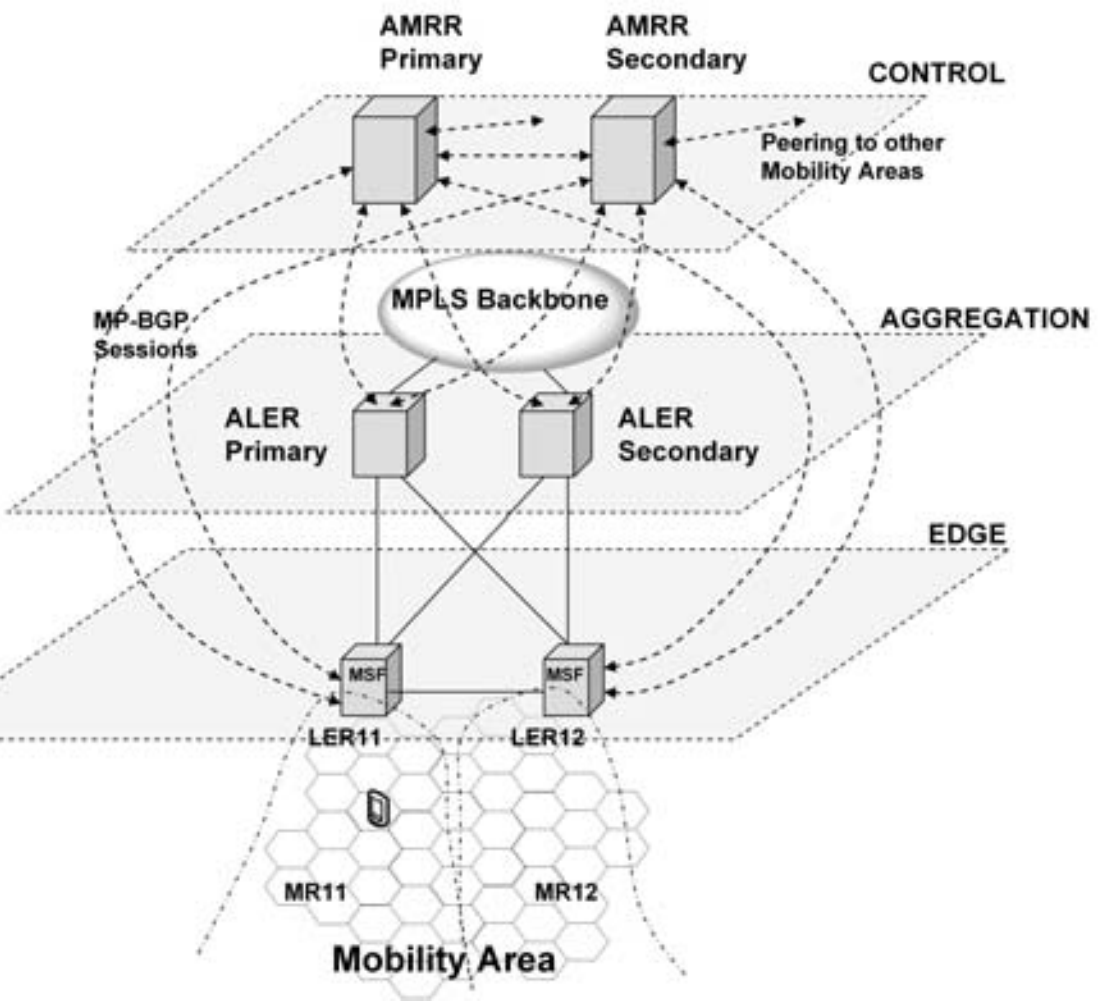

Figure 17. High Availability configuration for Mobility Area.

If a mobile device registers with the MSF in LER11, the internal Mobility Binding update is sent to both AMRR nodes, and this update is reflected to both ALER nodes. The ALER nodes allocate the Local Mobility Labels and update both AMRR nodes using the external update. The AMRR nodes now have the Mobility Binding information for the mobile device represented by two ALER nodes for the requests from outside the area.

At this point the AMRR nodes may service the requests from outside the area by replying with the Mobility Bindings that contain the Router IDs of both ALER nodes following a certain distribution algorithm (for example, some form of a round-robin Router ID selection). This allows for load distribution among the two ALER nodes.

If one AMRR node fails the peering sessions with the corresponding LER, ALER and other AMRR nodes will also fail and will be detected by the peering nodes. The remaining AMRR will have all the required information and is still able to serve the area.

If one of the ALER nodes fails, the corresponding failure of the MP-BGP peering sessions will be detected by both AMRR nodes. The AMRR nodes in turn may issue a special blanket update message containing the Router ID of the remaining ALER node to all peering AMRR nodes in other areas to use that Router ID in all Mobility Bindings received from the area. Once the ALER nodes in other areas are updated the network wide fail-over to the remaining ALER node in the local Mobility Area may take place.

It is important to note that the ALER failure update message does not need to reach all LER nodes in all areas, just the ALER nodes need to be updated. The redirection of traffic is then achieved by using the segmented LSPs that terminate at the surviving ALER node.

### C. *Scope of Mobility Labels*

The Mobility Labels are allocated in two places in the H-MLBN – the edge level LER nodes and the aggregation level ALER nodes.

The LERs allocate Mobility Labels upon registering a mobile device and use internal updates to communicate the associated bindings to the AMRR. If there is a CN in the same Mobility Area that wishes to communicate with a registered MN, the LER serving the CN may issue a Mobility Binding request to the AMRR. The AMRR in this case will reply with the Mobility Binding containing the Router ID of the originating LER as the value of the “Origin MP-BGP Next-Hop”. Thus the LSP carrying the traffic between MN and CN will terminate at the two edge level LER nodes. If this LSP needs to pass through the ALER node (due to the physical connectivity requirements) the ALER node will not need to examine the Mobility Label and switch the packets using only the top labels associated with the Router IDs of the two edge LER nodes.

Thus the Mobility Labels allocated by the edge level LERs are only scoped to the local Mobility Area. Their uniqueness is guaranteed by the fact that they are always associated with the unique Router IDs of the LER nodes and the corresponding top level labels. This also means that these labels are locally significant to the edge LER nodes. In other words, if two LERs assigned the same value of the Mobility Label to two different MNs and these MNs need to communicate with each other – this situation does not present a conflict. The Mobility Label and the top label associated with the far end Router ID will be used to send traffic, and the Mobility Label received in the incoming packet will be used to lookup the IP address of the locally registered MN and to take any further required local processing action.

The ALERs allocate a Local Mobility Label when an internal or external Mobility Binding update is received. This Local Mobility Label is used to represent the internal MNs to other Mobility Areas as well as to represent the external MNs to the local area. Thus the Local Mobility Labels are only significant within the ALER nodes and their uniqueness is controlled by the label allocation process within the ALER. These labels are only scoped to the path between the two ALER nodes from different Mobility Areas.

In other words, if two edge LER nodes within an area allocated the same Mobility Label value to two different MNs and CNs from outside the area need to communicate with the MNs via the ALER – this situation does not present a conflict. This is because the ALER allocates two different Local Mobility Labels to the Mobility Bindings associated with the MNs and keeps the label trail in its FIB linking the Current Mobility Label allocated by the edge LER to the Local Mobility Label. This Local Mobility Label along with the ALER Router ID (as opposed to the edge LER Router ID) is used in the Mobility Bindings sent to other areas. As MNs move within the area, the value of the Current Mobility Label is updated using internal updates and the value of the Local Mobility Label does not change. This results in no need to update the rest of the network when MNs move within the area.

The situation is reverse for the case when external MNs are represented in the local area by the Local Mobility Label assigned by the ALER and communicated to the edge LERs using the internal update via the AMRR node. As the Mobility Labels for the MN change (due to the movements between the areas), the ALER updates the label trail and no further internal updates to the LER nodes are required – the existing Local Mobility Label is used.

It is important to note that in some cases the scope of the Mobility Labels assigned by the ALER nodes may change and may extend to certain edge LER nodes. An example of this is when an edge LER node is or connects to the service provider's Internet peering point. In this case the ALER nodes in other areas may communicate their Local Mobility Labels representing the MNs within their areas directly to such LER node. The uniqueness of labels is guaranteed by the fact that they are always associated with the unique Router IDs of different ALER nodes and the corresponding top labels to reach these Router IDs.

### D. *Scope of Mobility LSPs*

The segmented LSP model (Fig. 7) avoids the requirement to update all edge level LER nodes in the network when MNs move within the Mobility Areas or between the Mobility Areas. The scope of the Mobility LSP is controlled by the Router ID used in the corresponding Mobility Binding as the value of the "Origin MP-BGP Next-Hop". In the segmented model the Mobility LSPs are always scoped to between the two edge LER nodes for intra-area traffic. For inter-area traffic the path is segmented and the Mobility LSP segments are scoped to between the near-end edge LER and the near-end ALER, between the two ALERs and between the far-end ALER and the far-end edge LER.

In cases where the Mobility Label assigned by an ALER node extends directly to an edge LER node (for example, the Internet peering point) in a different area, the LSP scope is also different. In this case the Mobility LSP extends from the edge LER node via the near-end ALER node (without the termination and the corresponding Mobility Label lookup) to the far-end ALER node where the segment is terminated and the new segment associated with the current Mobility Label is originated.

It is important to point out that the Mobility LSP scope is flexible and may be controlled by the logic in the AMRR function.

### E. *Mobile Node Multi-Homing and Traffic Continuity*

The preservation of traffic continuity during the mobile node movements is a very important requirement and is achieved by coordinating the layer 2 hand-off with the layer 3 hand-off processes. The layer 2 hand off takes place when a mobile node transitions from one RAN base station to another and is usually based on the radio signal level that the MN receives from different base stations.

Regardless of the type of the layer 2 hand-off (hard or soft) the important point is that there is a period of time during which the MN has connectivity to two different RAN base stations. If the two base stations in question are associated with the same layer 3 domain or sub-net (such as the same layer 3 logical interface of the MSF) by way of the wire-line layer 2 grooming network configuration, the layer 2 hand-off is transparent to the network layer and the traffic continuity is only a function of the RAN hand-off process.

If the layer 2 hand-off results in the MN moving to a different layer 3 domain, the traffic continuity also depends on the layer 3 hand-off process. The coordination between the layer 2 and layer 3 hand-off processes may be explicit (based on the cross layer signaling) or implicit (based on the built-in mobility management scheme properties).

There is also some level of coordination within the MN itself. For example, when the MN detects a stronger radio signal or in the process of switching the RAN types (e.g., cellular to Wi-Fi), the hand-off logic may invoke the new layer 3 Discovery and Registration on the new RAN while the old RAN connection is maintained (radio signal quality permitting).

Thus in the context of H-MLBN, traffic continuity and MN multi-homing may be considered for the three hand-off modes (MSF-Local, Inter-MSF Intra-Area and Inter-MSF Inter-Area).

*During the MSF-Local hand-off* (Fig. 11), while the MN is multi-homed to two RAN base stations, traffic continuity is controlled by the MSF layer 3 tracking function. In its simplest form the continuity logic may wait until the mobile node sourced IP packets start arriving on the new layer 3 logical interface (similar to Cellular IP [8]) and then update the local registration record for the MN with the new interface ID. A more elaborate scheme may involve packet replication toward the MN across the two layer 3 logical interfaces for some period of time (maintaining a registration record with two layer 3 interface IDs until no activity is detected by the tracking function on the old interface ID).

*During the Inter-MSF Intra-Area hand-off* (Fig. 12, 13), the MN is multi-homed to two RAN base stations that are under control of two different MSFs. The traffic continuity depends on the intra-area Mobility Binding update process. The basic form of traffic continuity involves the MN detecting a new RAN and initiating the Discovery and Registration with the new MSF while maintaining connectivity to the old MSF. While the intra-area update is taking place (AMRR update with the current Mobility Label by the new LER and the ALER update) the traffic to the MN may be delivered via the old Mobility LSP segment (from ALER to the old LER). Once the ALER has been updated with the new Current Mobility Label and the FIB label trail record is updated, the traffic may switch to the new Mobility LSP segment (New LSP in Fig. 12). A more sophisticated scheme may involve temporary packet replication at the ALER over the two existing Mobility LSP segments.

*During the Inter-MSF Inter-Area hand-off* (Fig. 14, 15), maintaining the traffic continuity is most difficult and depends on the inter-area update process. As in the above discussion, the MN detects a new RAN and initiates the Discovery and Registration with the new MSF while maintaining connectivity to the old MSF. The inter-area update process may be considered as a two-stage process.

The first stage involves the internal AMRR update by the new LER, the combined LRL request and the Mobility Binding update from the new AMRR to the old AMRR (based on the value of the Area ID received from the MN) and the external update of the old ALER by the old AMRR.

The second stage consists of the Mobility Binding update from the new serving AMRR (AMRR2 in Fig. 14) to the AMRR identified in the received LRL (AMRR3 in Fig. 14) and the external update of the ALER by the AMRR in the area serving the CN (ALER3 in Fig. 14).

Until the completion of the first stage the traffic to the MN may be delivered by the original old Mobility LSP (LER-ALER3-ALER1-LER in Fig. 14). Immediately after the completion of the first stage, right after the old ALER (ALER1 in Fig. 14) updates its FIB label trail, the traffic to the MN may be delivered using the interim Mobility LSP (shown in Fig. 15 – LER-ALER3-ALER1-ALER2-LER). And after the completion of the second stage, when ALER3 updates its FIB label trail, the traffic may switch to the new Mobility LSP (LER-ALER3-ALER2-LER).

Clearly the most vulnerable time period for the Inter-MSF Inter-Area hand-off is between the start of the layer 2 hand-off and the moment when the first stage of the inter-area update is completed. Compared to the Intra-Area hand-off the first stage involves an extra inter-AMRR update. In addition during the time period between the completion of the first stage and the completion of the second stage the traffic will be temporarily following a sub-optimal forwarding path.

*F. Role of Area ID*

The purpose of the Area ID is to track the last Mobility Area visited by a mobile node and to identify which Mobility Areas requested Mobility Binding information for the mobile node during the lifetime of the Mobility Binding. The last visited Area ID is communicated to the MN by the LER during the registration process with the MSF. Area ID may be stored in the MN's memory and included into the MSF Discovery and Registration messaging.

Initially in a start-up state, MN uses the Area ID value of 0 in the MSF registration messaging. If the LER receives the Area ID value of 0 from the MN, the serving LER updates the MN with its own Area ID and uses its own Area ID during the internal update of the AMRR. On all subsequent moves, the MN should be using the Area ID value other then 0. If the serving LER receives the Area ID value that is not equal to 0 from the MN, the LER passes the received Area ID value unchanged to the AMRR node during the internal update.

When the AMRR node receives an internal Mobility Binding update from an LER node it must check the Area ID value in the received information. If the Area ID equals its own Area ID it reflects the update to the ALER node and takes no further action. Otherwise, the AMRR sends the LRL request to the AMRR identified by the originally received value of the Area ID, re-writes the received Area ID with its own and reflects the internal update to the ALER. The Area ID check by the AMRR is only performed on the internal Mobility Binding updates. When an external update is received by an AMRR node the Area ID in the Mobility Binding is simply stored in the MN binding record.

The LRL received in the reply from another AMRR node contains a list of Area IDs from which the requests for the Mobility Binding information for the MN in question were received. The AMRR node then sends unsolicited Mobility Binding updates to the corresponding AMRR nodes with the listed Area IDs (as discussed in the Inter-Area hand-off).

If an AMRR node receives a Mobility Binding request from another peer AMRR node it should send a positive reply only for the locally stored Mobility Bindings that have their Area ID values equal to the Area ID of the responding AMRR node. Otherwise the AMRR should send a negative reply to the received request. The Area ID (AID) processing is summarized in Fig. 18.

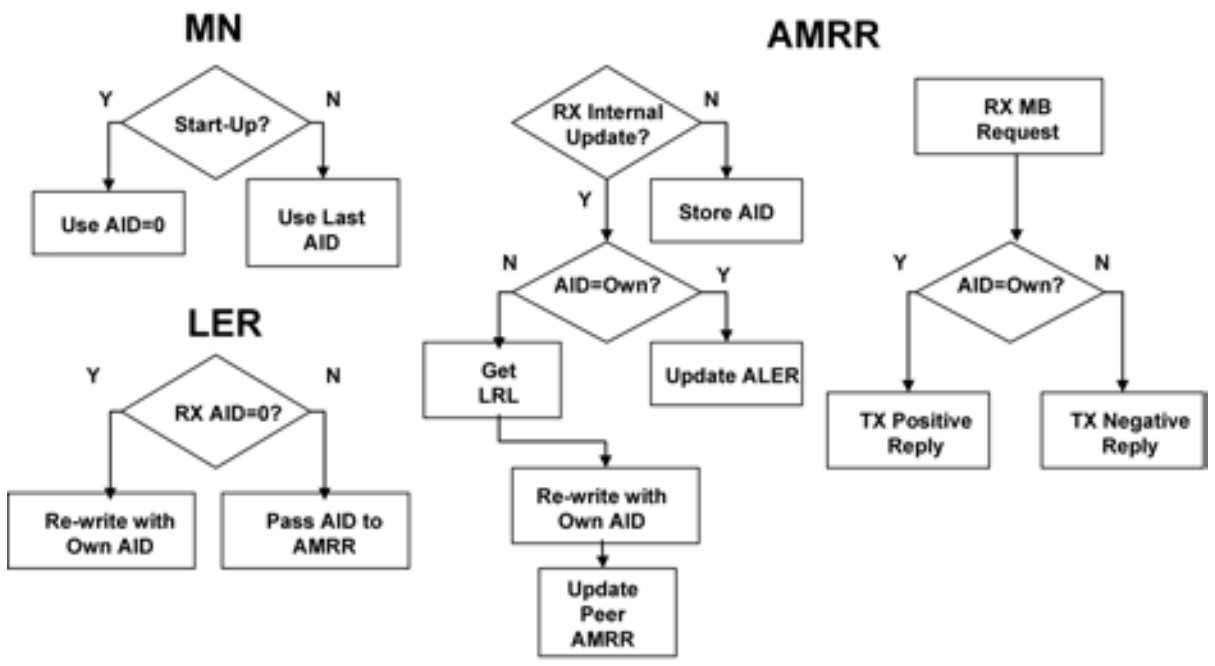

Figure 18. Area ID Processing logic.

*G. Mobility Binding Lifetime and Transient Conditions*

Clearly, Mobility Bindings should not be allowed to exist in the network forever and must have an expiration time associated with them. This time interval (referred to as a *Lifetime*) may be carried explicitly with each Mobility Binding or set as a network wide parameter for all bindings. In any case the Lifetime starts counting down from the moment the binding information has been received by a network node (LER, AMRR, ALER). Once the Lifetime expires the binding is silently removed from each node's database.

Given that the Mobility Bindings have a finite Lifetime, there are some specific transient conditions that may lead to potentially significant interruption in traffic delivery to mobile nodes. One such condition is when during an active Mobility Binding Lifetime a mobile device that has previously been registered in a given Mobility Region resets, and shortly thereafter a mobile user moves into another Mobility Area.

In this case the old local AMRR and ALER nodes still count down the Lifetime for the MN's binding marked with their own Area ID (thinking that the MN still exists in the local area). The AMRR, ALER and LER nodes in other areas that the MN has been communicating with, also have the "stale" binding and the associated label stack leading to the old local ALER node.

When the MN goes through the start-up mode in the new region and new area, it must use the Area ID value of 0. The new local LER will re-write this Area ID with the new Area ID and send an internal update to the new AMRR and so on (as in the described start-up sequence). Since the new AMRR may not have any old active bindings associated with the MN, it stores the newly received binding information ready to reply to the requests from internal and external nodes. The issue is that due to the fact that a stale binding may still exist in other external LER and ALER nodes pointing to the old local ALER for the MN, and until this binding expires the traffic to the MN will be directed to a wrong place.

To solve this problem the Mobility Binding Withdrawal mechanism is required. Clearly to be efficient the withdrawal should not be based on flooding the entire network with the MP-BGP Update (carried in the MP-UNREACH-NLRI attribute with the encoded Mobility Binding that is being withdrawn [2]).

In order to facilitate the scalable distribution of withdrawal messages, the LRL (Last Requestor List) processing is enhanced to include two types of LRLs at the AMRR level: External LRL (eLRL) and Internal LRL (iLRL).

The eLRL is the same as described earlier (LRL) and consists of the Area IDs from which the requests for the Mobility Binding in question have been received by an AMRR node from external AMRR nodes during the Lifetime of the binding.

The iLRL is also maintained by an AMRR node and is a list of the Router IDs of all LER nodes internal to an area which originated Mobility Binding requests for the binding in question.

The Mobility Binding Withdrawal sequence follows the steps below:

1. When an LER node detects loss of communication with a registered MN (based on the keepalive mechanism) it sends an internal Mobility Binding Withdrawal update to the local AMRR using the MP-UNREACH-NLRI encoding [2]. The LER clears the Mobility Binding and the registration record from its memory.
2. The local AMRR receives the withdrawal message and looks up the Mobility Binding record. If at this time the AMRR finds an existing Mobility Binding associated with a Router ID of another LER and different from the LER that originated the withdrawal, the AMRR ignores the received withdrawal and takes no further action.
3. Otherwise, if the Mobility Binding exists and the "Origin MP-BGP Next-Hop" value matches the Router ID of the LER originating the withdrawal, the AMRR locates the iLRL and reflects the withdrawal update to all LERs whose Router IDs have been found in the iLRL. In addition, the AMRR locates the eLRL and forwards the withdrawal update to all AMRR nodes whose Area IDs have been found in the eLRL. AMRR removes the binding.

4. Each AMRR node that receives the Mobility Binding withdrawal update from another AMRR looks up the iLRL and forwards the update to all LERs with the Router IDs found in the list. The AMRR then clears the Mobility Binding.
5. Each LER that receives the Mobility Binding withdrawal update clears the Mobility Binding and the associated label stack from its memory.

The result is that the network may be updated in a scalable manner before the Mobility Binding Lifetime expires. Step 2 in the above sequence is designed in order to prevent the unnecessary inter-area withdrawals (it implies that the MN has re-registered with another LER local to the same or another area and the Intra-Area or Inter-Area hand-off sequence has been executed before the withdrawal update has been received).

Clearly there has to be a defined relationship among the following time intervals: Mobility Binding Lifetime (*L*), Registration Dead Time (*D*) and the Expected Re-registration Time (*R*). In general this relationship should be expected to follow: $L >> D >> R$.

The Mobility Binding Withdrawal process state diagram is shown in Fig. 19.

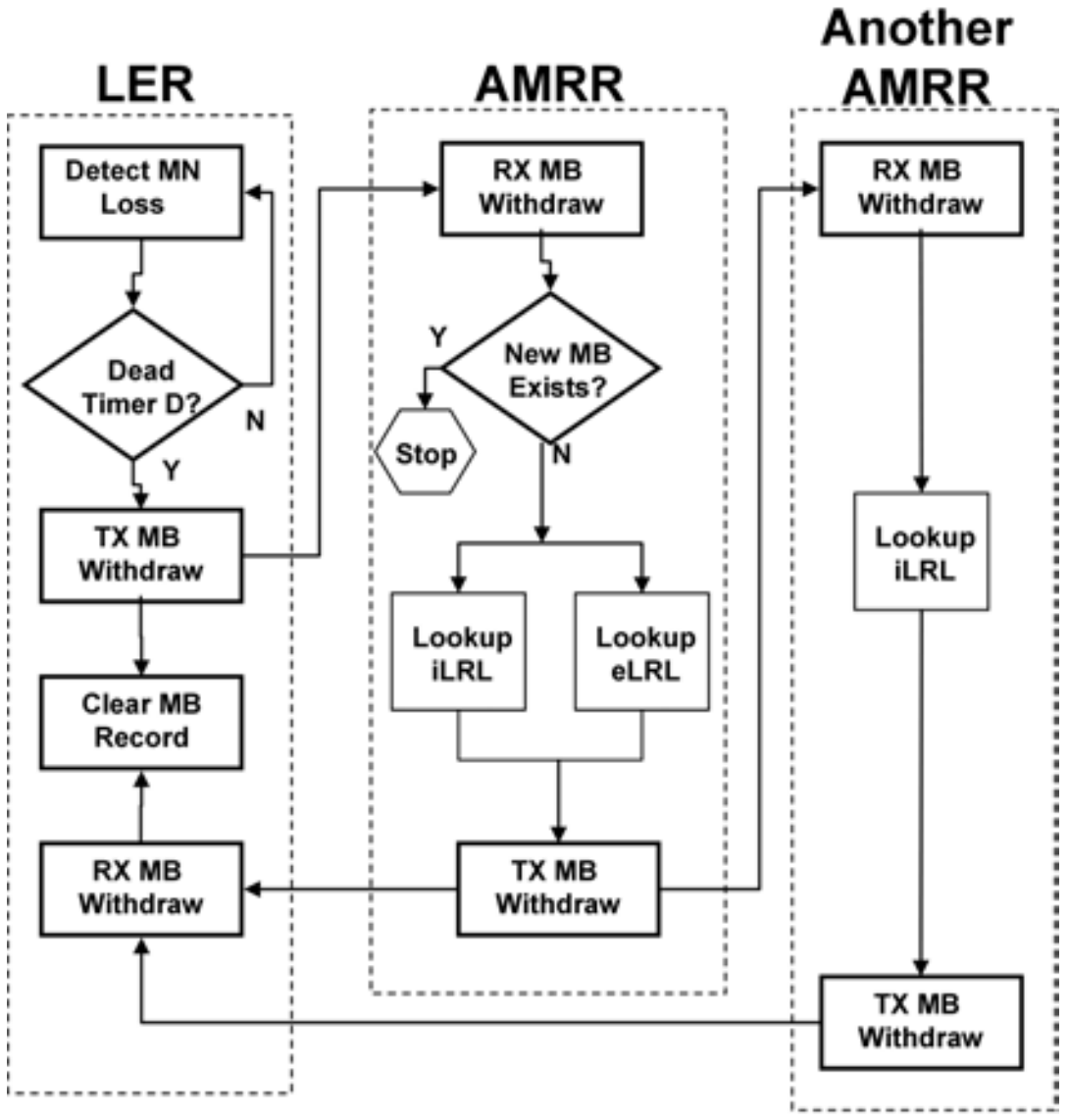

Figure 19. Mobility Binding Withdrawal Process. MB – Mobility Binding, D – Registration Dead Time, iLRL – Internal Last Requestor List, eLRL – External LRL.

## VI. Other Applications for Mobility Labels

Mobility Label is a critical component in building the network hierarchy necessary for the increased scalability of the overall solution. This section describes other uses for the Mobility Label that involve processing manipulations at the serving MSF on the received MPLS packet destined towards a mobile node.

### A. RAN Differentiation

In order to speed up traffic processing at the MPLS LER nodes facing the RAN the Mobility Label assigned by the LER node and used in the received packet's MPLS label stack in the packets destined to the mobile node may be assigned such that the layer 2 properties of the packet to be forwarded to a particular RAN type are easily identified and the associated protocol headers are pre-constructed and stored in the LER memory along with the mobile node's registration record.

For, example if an LER is serving different RAN types such as CDMA, GSM, Wi-Max or Wi-Fi, each of the served RAN technologies may require different types of layer 2 packet encapsulations and addressing. At the time when a mobile node registers with the serving MSF, a Mobility Label is assigned and associated with the registration record including at a minimum the mobile node's IP address, layer 2 address, Area ID and the associated layer 3 logical interface of the MSF. In addition to this information the LER may pre-construct a RAN-specific layer 2 protocol header, including the encapsulation and addressing information, and attach it to the record.

When a MPLS packet arrives at the LER, the LER pops the top label, reads the Mobility Label and proceeds directly to building the RAN-specific layer 2 packet followed by the IP header and payload using the information stored in the registration record and identified by a fixed-size locally significant Mobility Label.

Without the pre-construction of the layer 2 information an LER would have to lookup all the components of the packet separately taking more time to switch the packet to the RAN. The local processing of the Mobility Label for the RAN differentiation offers a simple and effective way of reducing the packet processing time.

### B. Normalized IP Address Lookup for Mobile Nodes

The MLBN is capable of supporting the mobile devices addressed using IPv4 or IPv6. IPv4 and IPv6 addresses are different in length (32 bits for IPv4 and 128 bits for IPv6) and are usually stored in different memory tables at the LERs. Clearly the IP address lookup times depend on the size of the address field and the size of the storing database. The presence of the fixed-length (20 bits) locally significant and unique Mobility Label allows to normalize the IP address search time (lookup time) at the MPLS LER serving the mixed IPv4/IPv6 mobile environment.

When a MPLS packet arrives at the LER, the LER uses the 20-bit Mobility Label to locate the IPv4 or IPv6 address of the mobile node in the associated registration record thus normalizing the processing time for both protocol stacks.

### C. Virtualization

The presence of Mobility Labels and the corresponding MPLS label stack allows utilization of the same MLBN infrastructure not only for providing efficient support for various access RAN types but also to enable virtualization of services.

As an example consider a MLBN providing services to multiple wireless carriers using the same MPLS network. Each carrier may have multiple RAN types and may be using unique or overlapping (private) IP address ranges for their mobile devices. In this case Mobility Labels (or even a special label stack – including the Carrier Label and the Mobility Label) may be used to differentiate among the multiple carriers and their RANs. This is similar to the layer 3 MPLS VPN service [4] and requires the corresponding support in MP-BGP for the address space differentiation and information partitioning (such as Route Distinguishers and Route Targets).

### D. Inter-Carrier Roaming

This use of Mobility Labels involves a scenario where a mobile device assigned a carrier-specific unique IPv4 or IPv6 address moves from the MLBN operated by one carrier to the MLBN operated by another carrier. If the mobile device is in an active communication session and needs to retain its fixed IP address a Mobility Label based interconnect between the service providers may enable this type of roaming.

In this architecture roaming between the service providers is supported by establishing the mobility enabled MP-BGP peering points between the provider networks. These peering points may be established between the special types of nodes referred to as the MLBN Border Edge Router (MBER). These nodes perform a function similar to the ALER nodes and peer to their local AMRR nodes.

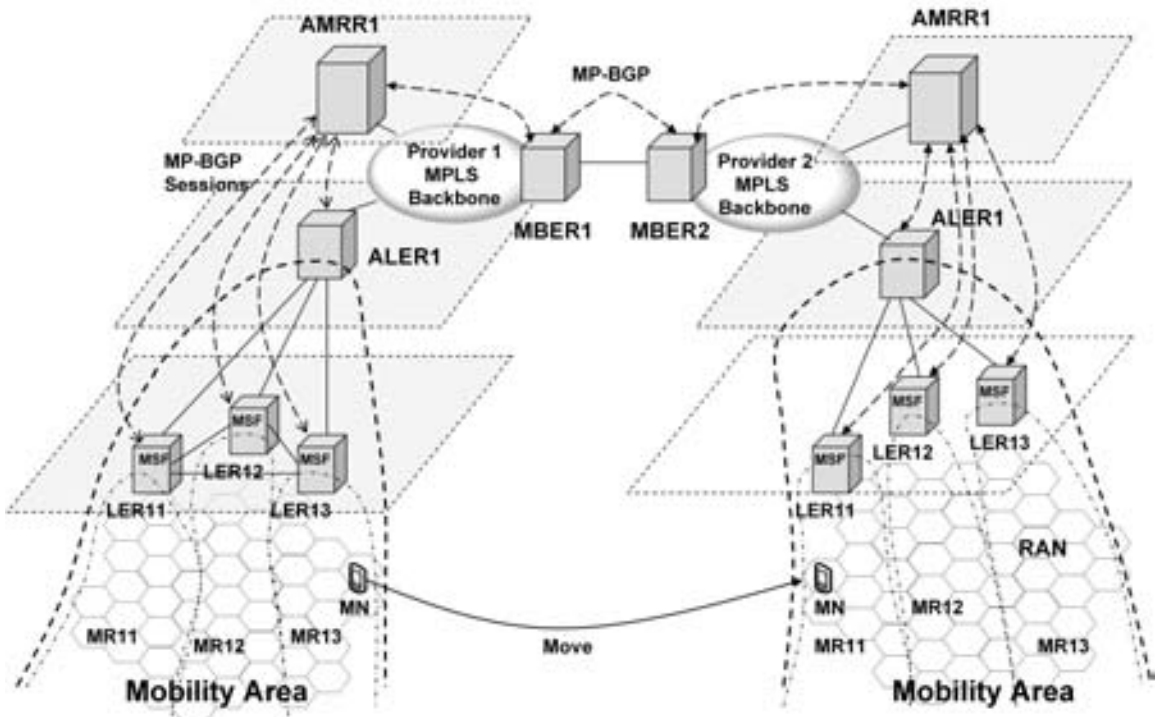

Figure 20. Inter-Carrier Roaming with H-MLBN

Consider Fig. 20, where two service providers are connected using the corresponding MBER nodes. The MBERs are peered with each other and with their local AMRR nodes. When a mobile node roams from a RAN served by one provider to a RAN served by another provider, the MN initiates the layer 2 hand-off during which it identifies the move to another service provider. The MN perform the MSF Discovery and Registration as in the "start-up condition" and registers with the MSF in LER11 of provider 2. The provider's LER updates the AMRR using the inter-carrier update type (indicated by the Update Type field in the Mobility Binding) and the AMRR reflects the update to the ALER. ALER allocates the Local Mobility Label and updates the MBER node via the corresponding AMRR (or optionally directly – not shown in Fig. 20). The MBER2 node stores the update, allocates its own Local Mobility Label and updates its peer MBER1 using its own Router ID as the "Origin MP-BGP Next Hop" in the Mobility Binding. The MBER1 then updates its own AMRR which in turn updates the ALER node using MBER1 Router ID. Traffic delivery to the MN follows the segmented Mobility LSP: ALER1_Provider1-MBER1-MBER2-ALER1_Provider2.

The segments of the LSP are terminated at each node and the next segment is identified by the value of the Local Mobility Label. When the MN moves within the provider 2 MLBN, the MSF-Local, Intra-Area and Inter-Area hand-off procedures are performed and provider 1 does not need to be updated. It is important to note that in order to ensure optimal traffic routing the peering MLBN providers must identify to each other their respective mobility address spaces.

## VII. Benefits of the MLBN approach

This section illustrates performance benefits resulting from the proposed mobility management approach. Although there are various benefits that may be attributed to this solution (see Section I) the main user facing advantage lies in the fact that it eliminates suboptimal routing (triangular routing and bidirectional tunneling) that is inherent in the mobility management schemes based on Mobile IPv4/v6 (see section II). To illustrate this we estimate the user facing penalty related to triangular routing by determining the average difference in the number of hops that a transmitted packet will experience if it is routed using triangular path compared to the optimal path, when a mobile node (MN) moves around the network and communicates with a fixed correspondent node (CN) or another moving MN This illustration assumes the use of reverse tunneling which is the most common choice in practical Mobile IP deployments.

We define optimal distance $\mathbb{k}^{o}$ as the distance between MN and CN expressed as a number of router hops along the optimal routing path. Triangular distance $\mathbb{k}^{t}$ is defined as the hop distance between MN and CN along the path from CN to the Home Agent (HA) and from HA to MN using optimal routing on each segment. A router hop is counted when a packet crosses a router. We assume a general network topology represented by a connected graph $G(V, E)$, where a mobility region (MR) is associated with each leaf vertex of $G$, and the maximum network diameter is $\mathbb{k}$ router hops.. An example of this topology is shown in Fig. 2.

We then consider a mobile node moving between the MRs and spending an average time of $1/\mu$ seconds in each region. We assume that the region dwelling time $t_{mr}$ is exponentially distributed with parameter $\mu$. We do not impose any specific restrictions on the pattern of movements of the mobile node, except to say that the probability of transitioning from one MR to another is $p$ (one example of a reasonable mobility pattern is to allow the mobile node to transition only between the adjacent mobility regions with probability $p = 1/2$).

We select one leaf vertex $v_i$ at random and designate it as the CN location, we then select another leaf vertex $v_j$ and designate it as the HA location. We move the MN through all of the leaf vertices of $G$ (including $v_i$ and $v_j$ ) and at every move of the MN determine the value of $d_i = \mathbb{k}^t - \mathbb{k}^o, \quad i = 0, ..., d_m$, where $d_m = \max(\mathbb{k}^t - \mathbb{k}^o)$.

Clearly, $d_i$ represents a difference in the number of router hops (vertices) in the communication path between MN and CN using triangular routing and direct routing. If this procedure is repeated for all possible locations for CN and HA, then a process $\{X(t), t \geq 0\}$ could be formed with states $d_i, \quad i = 0, ..., d_m$. Given a general topology of graph $G$, there may be more than one leaf vertex in $G$ with the same degree of difference $d_j$ relative to a given position of CN and HA. At the same time there will always be at least one vertex with $d_j = 0$ and at least one with $d_j = d_m$.

It may also be reasonably assumed that the MN may transition to a MR represented by a vertex with any degree of difference $0 \leq j \leq d_m$ at any step with equal probability $q = 1/d_m$. In addition, the dwelling time $t_{d_j}$ of process $X(t)$ in a state $d_j$ is also exponentially distributed. To see this, one can think that it takes $N$ steps of the MN for the process $X(t)$ to remain in a given state $d_i$ before moving to another state $d_j$, where $N$ is geometrically distributed with parameter $p$. And since each step $k = 1, ..., N$ of MN takes an exponentially distributed time $t_{mr}$ the total time spent by process $X(t)$ in any given state $i$ is $t_{d_i} = \sum_{k=1}^{N} t_{mr_k}$, which is a geometrically compounded sum of i.i.d exponential random variables, meaning that $t_{d_i}$ is also exponentially distributed with parameter $\eta = p\mu$.

Thus $\{X(t), t \geq 0\}$ forms a Continuous Time Markov Chain (CTMC) with the state transition-rate diagram shown below.

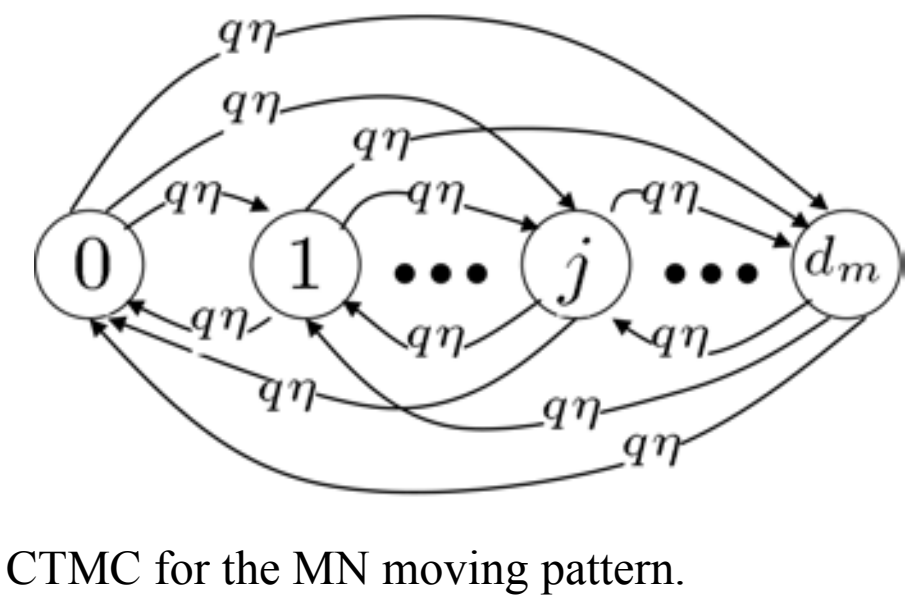

Figure 21. CTMC for the MN moving pattern.

Using global balance equations, the stationary distribution can be computed as follows:

$$\pi_0 d_m q\eta = \sum_{j=1}^{d_m} \pi_j q\eta = q\eta(1 - \pi_0) \implies \pi_0 = \frac{q}{1+q} \qquad (1)$$

and for $0 \leq j \leq d_m$:

$$\pi_j = \frac{q}{1+q} = \frac{1}{(d_m + 1)} \qquad (2)$$

Thus, we can find the expected value of $X(t)$ as:

$$E[X(t)] = \sum_{j=0}^{d_m} j\pi_j = \frac{1}{(d_m+1)} \sum_{j=0}^{d_m} j = \frac{d_m(d_m+1)}{2(d_m+1)} = \frac{d_m}{2} \qquad (3)$$

The value of $d_m$ may vary depending on the actual count of router hops along the optimal path. However, the following upper and lower bounding of $d_m$ may be considered:

$$0 \leq d_m \leq (\max(\mathbb{k}^t) - \min(\mathbb{k}^o)) \qquad (4)$$

The right hand side of (4) corresponds to a pessimistic case where the upper bound of $d_m$ may be considered with $\min(\mathbb{k}^o) = 1$, and the left hand side of (4) corresponds to an optimistic case where there is no difference between the two paths. For the purposes of this illustration we choose to use the upper bound of $d_m: \quad d_m^u = \max(\mathbb{k}^t) - 1$. We consider two scenarios when establishing the value of $d_m^u$: i) a moving MN communicates with a fixed CN, and ii) two moving MNs communicate with each other. In the first scenario, the worst case for triangular routing is $\mathbb{k}^t = 2(\mathbb{k} - 1) + 1 = 2\mathbb{k} - 1$ and $d_m^u = 2(\mathbb{k} - 1)$. In the second scenario, the worst case for triangular routing is $\mathbb{k}^t = 3(\mathbb{k} - 1) + 2 = 3\mathbb{k} - 1$, and therefore $d_m^u = 3\mathbb{k} - 2$.

The expected value of $X(t)$ may be written as:

$$\overline{d} = E[X(t)] = \begin{cases} (\mathbb{k} - 1), & \text{for fixed-to-mobile case} \\ (3\mathbb{k} - 2)/2, & \text{for mobile-to-mobile case} \end{cases} \qquad (5)$$

Expression (5) may be interpreted as follows: given a general network topology with the maximum network diameter of $\mathbb{k}$ router hops and a general mobile node moving pattern with exponentially distributed dwelling times, the average difference in the number of router hops between the communicating nodes using triangular routing versus optimal routing increases linearly with the maximum network diameter. It is also important to note that this difference increases 1.5 times on average when two mobile nodes are communicating with each other compared to the case where a fixed CN communicates with a moving MN.

In order to translate this result into the cost to the user we associate delay and packet loss with each router hop. We assume that delay $\delta_i$ is normally distributed with mean $\alpha$ and standard deviation $\zeta$ ($\delta_i \backsim G(\alpha, \zeta^2)$), and that at each hop there is a probability $p_l$ of loosing a packet.

Thus the cumulative delay penalty $C_\delta$ is also normally distributed with mean $\alpha\overline{d}$ and variance $\zeta^2\overline{d}$, and the probability of losing a packet due to the extra traveled hops is $C_l = 1 - (1 - p_l)^{\overline{d}}$. This is illustrated in Fig.22 for $\alpha = 5$ milliseconds, $\zeta = 2$ milliseconds and $p_l = 0.005$.

As can be seen from Fig. 22, given the listed parameters, with a maximum network diameter of 10 hops the increase in one-way delay due to triangular routing for mobile-to-mobile communication is approximately between 60 and 80 milliseconds (based on a single standard deviation interval), along with the corresponding decrease in probability of successful packet delivery of nearly 7%.

In summary the user facing penalty for triangular routing is increased delay and jitter (delay variation), accompanied by increased loss probability that grow linearly with the maximum network diameter.

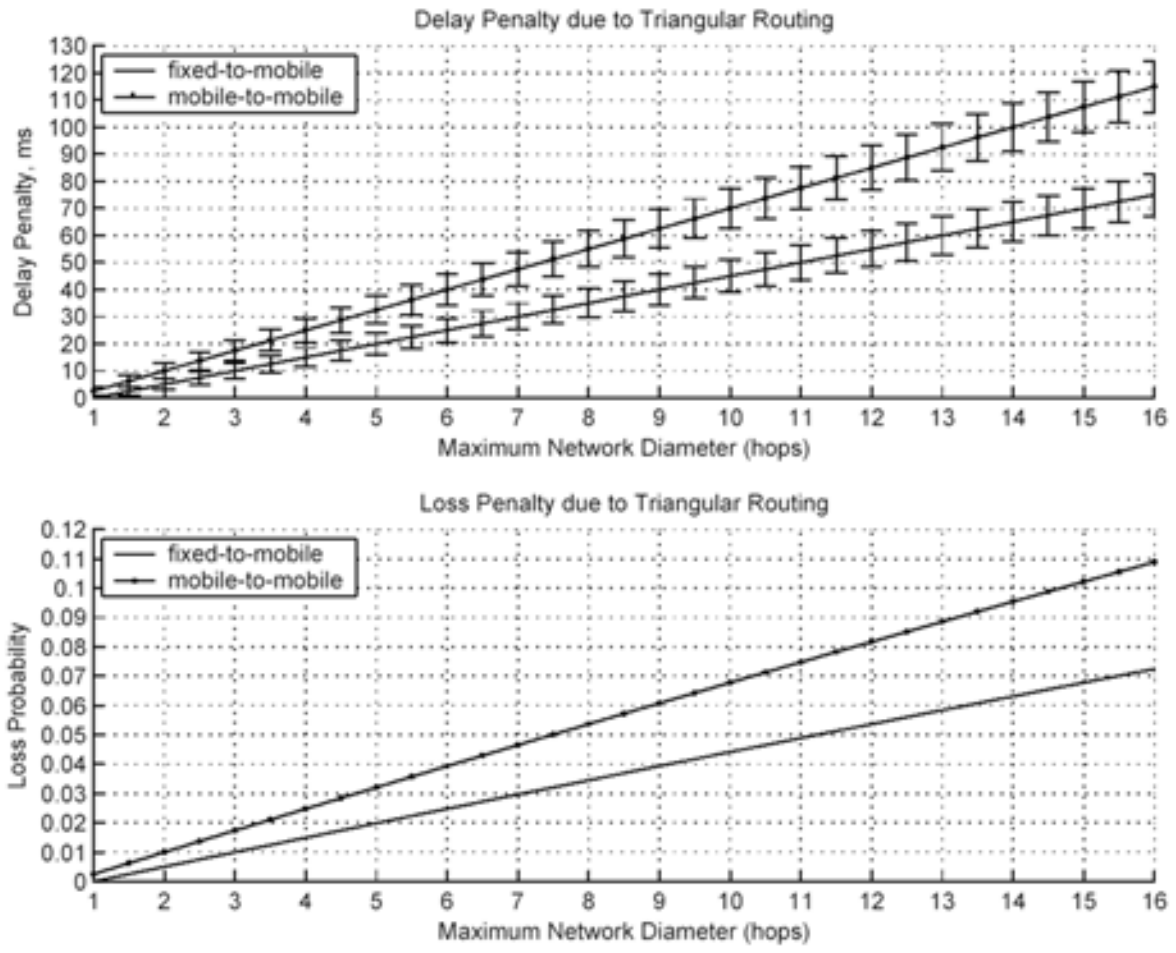

Fig. 22. Delay (top) and Loss penalty due to triangular routing. Delay penalty is shown within one s.t.d from the mean.

## VIII. Conclusions

The Hierarchical Mobility Label Based Network (H-MLBN) represents a mobility management system in which mobility control plane is fully integrated with MPLS forwarding plane resulting in optimal traffic delivery to mobile devices.

H-MLBN provides support for macro- and micro-mobility for IPv4 and IPv6 mobile hosts and routers under a common MPLS-based control plane without a need for Mobile IP.

At the forwarding plane H-MLBN is designed to avoid explicit MPLS LSP creation, teardown and redirection following the movements of mobile devices. On the contrary H-MLBN relies on the use of pre-constructed full logical mesh of LSPs connecting all LERs and LSRs in the network and managed by the Label Distribution Protocol. The task of mapping mobile prefixes to the existing LSPs is performed by the H-MLBN mobility control plane using a modified MP-BGP protocol. This mapping however is not a simple IP prefix to LSP label mapping. Such a mapping approach would require that the IP addresses of mobile devices are looked up at every intermediate H-MLBN node thus resulting in a significant increase in a processing load and reduction in scalability.

A Mobility Label is introduced and associated with every mobile prefix and the Router ID of an H-MLBN node that is used to reach the mobile device. This Mobility Label is used as the second label in the MPLS label stack allowing intermediate H-MLBN nodes to avoid IP prefix lookups and just perform conventional MPLS level operations (pop, swap, push) while forwarding traffic to mobile devices over a set of LSP segments.

The control plane architecture of H-MLBN is constructed in a hierarchical manner providing support for micro-mobility. Micro-mobility is achieved through the support of MSF-Local and Intra-Area layer 3 hand-offs. MSF-Local hand-off is performed locally by an LER node and does not require any network updates and associated state changes. Intra-Area hand-off is performed within a Mobility Area by an interaction between LER, AMRR and ALER nodes, and does not require network updates and state changes outside of the Mobility Area.

The regionalized network architecture supported by specific control plane mechanisms such as mobile device to MSF-logical-interface association tracking, mobility area internal updates, on-demand mobility binding requests, Area ID processing and segmentation of mobility LSPs, allows to minimize the frequency of network-wide updates and eliminates the need to share the visitor information on all network nodes participating in mobility management.

The “fast versions” of MIPv4 [22] and MIPv6 [23] propose to use layer 2 triggers and additional layer 3 signaling to achieve fast handoff when MN moves from one access network to another. Since the use of layer 2 triggers to facilitate the layer 3 handoff is not specific to Mobile IP similar mechanisms may be introduced to H-MLBN in the future enhancements.

However, as described in [1] MLBN supports a concept of a Group Registration by which MN’s registration information is distributed to a pre-defined group of neighboring LER nodes and is maintained through the use of specially encoded MP-BGP updates [2]. In addition, a specific layer 2 and layer 3 virtual addressing scheme for the MSF logical interfaces in the LER nodes may be adopted allowing MNs to continue sending and receiving packets from the new MSF within the Registration Group immediately after or even during the layer 2 handoff thus minimizing the handoff latency and the associated service disruption.

Although existing Mobile IP based mobility management schemes allow for route optimization the specific mechanisms involved in these solutions are often difficult for practical implementation as they impose significant requirements on the fixed correspondent nodes and in turn generate sophisticated security schemes and requirements on mobile devices, home agents and fixed devices.

H-MLBN avoids these complexities and allows for optimal traffic delivery. The introduction of the hierarchy does not result in single points of failure and bottlenecks as the mobility related information and processing load may be flexibly distributed by the control plane.